\documentclass[11pt]{article}

\PassOptionsToPackage{nosort}{cite}
\PassOptionsToPackage{breaklinks}{hyperref}

\usepackage{amssymb}
\usepackage{amsfonts}
\usepackage{dsfont}
\usepackage{mathrsfs}
\usepackage{enumitem}
\usepackage[T1]{fontenc}
\usepackage{chet}
\usepackage{multirow}
\usepackage{longtable}
\usepackage[font=small,format=hang,labelfont={sf,bf}]{caption}
\usepackage{float}
\usepackage{subcaption}
\usepackage{pgfplots}

\usetikzlibrary{calc, external, intersections, arrows.meta, arrows, patterns, plotmarks, shapes.geometric, pgfplots.groupplots}

\pgfplotsset{width=10cm, compat=1.9,every non boxed y axis/.style={}}
\usepgfplotslibrary{colorbrewer}

\DeclareMathAlphabet{\mathdutchcal}{U}{dutchcal}{m}{n}
\SetMathAlphabet{\mathdutchcal}{bold}{U}{dutchcal}{b}{n}
\DeclareMathAlphabet{\mathdutchbcal}{U}{dutchcal}{b}{n}

\definecolor{darkblue}{rgb}{0.1,0.1,0.7}
\hypersetup{colorlinks,
           linkcolor={darkblue},
           citecolor={darkblue},
           urlcolor={darkblue},
           pdftitle={Multiscalar Critical Models with Localised Cubic Interactions},
           pdfdisplaydoctitle
}

\newcommand{\lsp}{\hspace{0.5pt}}
\newcommand{\lnsp}{\hspace{-0.5pt}}

\renewcommand{\geq}{\geqslant}
\renewcommand{\leq}{\leqslant}

\renewcommand{\vec}[1]{\mathbf{#1}}

\title{Multiscalar Critical Models with\\[7pt] Localised Cubic Interactions}

\author{Sabine Harribey,$^{
\lnsp a}$ William H.\ Pannell,$^{\lnsp b}$ and Andreas Stergiou$^{b,c}$\emails{\href{mailto:sabine.harribey@su.se}{sabine.harribey@su.se}, \href{mailto:william.pannell@kcl.ac.uk}{william.pannell@kcl.ac.uk}, \href{mailto:andreas.stergiou@kcl.ac.uk}{andreas.stergiou@kcl.ac.uk}}}

\affiliation{$^a$NORDITA, Stockholm University and KTH Royal Institute of Technology,\\[-3pt]
Hannes Alfv\'{e}ns v\"{a}g 12, SE-106 91 Stockholm, Sweden\\
$^b$Department of Mathematics, King's College London, Strand, London WC2R 2LS, United Kingdom\\
$^c$Theoretical Physics Department, CERN, 1211 Geneva 23, Switzerland}

\abstract{Interface localised interactions are studied for multiscalar universality classes accessible with the perturbative $\varepsilon$ expansion in $4-\varepsilon$ dimensions. The associated beta functions at one loop and partially at two loops are derived, and a wide variety of interface conformal field theories (CFTs) is found, even in cases where the bulk universality class is free or as simple as the Wilson--Fisher description of the $O(N)$ model. For up to three scalar fields in the bulk, interface fixed points are classified for all bulk universality classes encountered in this case. Numerical results are obtained for interface CFTs that exist for larger numbers of multiscalar fields. Our analytic and numerical results indicate the existence of a vast space of interface CFTs, much larger than the space of defect CFTs found for line and surface defect deformations of multiscalar models in $4-\varepsilon$ dimensions. In this vast space, stable interfaces found for free and $O(N)$ bulks belong to the $F_4$ family, with global symmetries $SO(3), SU(3), Sp(6)$ and $F_4$, realised with $N=5,8,16,24$ scalar fields, respectively.}

\date{July 2024}

\preprint{CERN-TH-2024-122}

\begin{document}

\maketitle

\toc

\section{Introduction}
The $O(N)$ vector model constitutes one of the most thoroughly studied class of conformal field theories (CFTs). While it does not appear to be exactly solvable by analytic means, there exist numerous methods to study it, such as the $\varepsilon$ expansion \cite{Wilson:1971dc}, the $1/N$ expansion \cite{Vasiliev:2003ev, Moshe:2003xn}, and the conformal bootstrap \cite{Polyakov:1974gs, Kos:2013tga, Poland:2018epd}. The $\varepsilon$ expansion is of particular importance, as it can also be used to study a variety of CFTs reached as endpoints of renormalisation group (RG) flows triggered by operators that break $O(N)$ symmetry. Indeed, generalising to quartic multiscalar models, it is possible to seek RG fixed points for different global symmetries, which are potentially relevant for second-order phase transitions observed in various physical systems. Extensive discussions of such fixed points can be found in \cite{Pelissetto:2000ek, Osborn:2017ucf, Rychkov:2018vya, Osborn:2020cnf}. In this work we will be interested in RG flows and CFTs that emerge when these fixed points are deformed by localised cubic interactions.

The study of systems with defects has a long history and renewed attention has been devoted to them recently. Defect deformations of CFTs give rise to new types of CFTs, called defect CFTs (dCFTs), that realise conformal symmetry in dimension given by that of the defect. This commonly includes dimension one, realised on line defects, but dimension two and the associated surface defects have also been widely discussed. Theoretical work outlining in detail the effects of the presence of defects and the observables that they give rise to can be found in \cite{McAvity:1995zd} for boundaries and interfaces and \cite{Billo:2016cpy} for more general defects. Such studies are of course motivated by the fact that defects describe numerous physical situations, such as the presence of impurities and localised perturbations, that can be realised experimentally; see for example \cite{sigl1986order, mailander1990near, burandt1993near, alvarado1982surface, PhysRevA.19.866, PhysRevB.40.4696, PhysRevB.58.12038}.

For the three-dimensional $O(N)$ model, numerous studies have been carried out for boundaries and defects. Such research dates back many years \cite{AJBray_1977, Ohno:1983lma, gompper1985conformal, McAvity:1995zd, Diehl:1996kd}, and more recently novel universality classes were determined for surface defects with quartic interactions in the bulk in exactly three dimensions \cite{Metlitski:2020cqy, Padayasi:2021sik, Toldin:2021kun, Krishnan:2023cff}.

Looking beyond the $O(N)$ model, RG properties of multiscalar models with defects open up an array of new research directions, as various fixed points can be found breaking the bulk symmetry in different ways. Besides their theoretical interest, such studies also have experimental applications. For example, in dimension three the cubic and Heisenberg model are famously hard to distinguish as their most easily accessible critical exponents are nearly identical; see e.g.\ \cite{Pelissetto:2000ek}. However, if the presence of a defect or interface breaks the bulk global symmetries in different ways, these two models could be distinguished without resorting to measurements of bulk critical exponents. This idea was explored in \cite{Pannell:2023pwz}, where a line defect perturbing a quartic multiscalar model was studied. In the case of an $O(N)$ bulk, a line defect can only break the bulk symmetry to $O(N-1)$. However, more complicated bulk symmetries, such as hypercubic or hypertetrahedral, allow for different patterns of symmetry breaking on a line defect. This fact, along with new observables that arise for a CFT in the presence of a defect, enable new computations and characterisations of bulk universality classes. Beyond line defects, \cite{Giombi:2023dqs, Trepanier:2023tvb, Raviv-Moshe:2023yvq} studied a surface defect perturbing the quartic $O(N)$ model and found that the bulk symmetry can only be broken to $O(p)\times O(N-p)$ with $p<N$ at leading order in the $\varepsilon$ expansion.

Here we will go one step further and study a defect with interactions cubic in the scalar field $\phi$. In this case we do not fix the dimension of the defect but its co-dimension. More specifically, we consider a co-dimension one defect, representing an interface. Our considerations generalise those of \cite{Harribey:2023xyv}, where the bulk symmetry was fixed to $O(N)$ and the interface symmetry to $O(N-1)$. Our deformation takes the general form
\begin{equation}\label{eq:gendef}
    S_{\text{interface}}=\int d^{\lsp 3-\varepsilon} \vec x\,
		\tfrac{1}{3!}h_{ijk} \lsp
		\phi_{i} \phi_{j} \phi_{k} \, , \qquad \, i,j,k=1,\dots,N \, ,
\end{equation}
where $\phi_i$ are scalar fields and $h_{ijk}$ are defect couplings. We add this deformation to $(4-\varepsilon)$-dimensional bulk CFTs with various global symmetry groups. We then study fixed points both analytically and numerically for different values of $N$. Generically, we find numerous fixed points that break the bulk global symmetry to a subgroup. The possible symmetry breaking patterns we obtain are far richer than for line and surface defects. We focus on unitary fixed points, as defined by real critical values of the defect couplings, although our methods can be used to find and analyse non-unitary fixed points as well.

In the case of a free bulk, operators cubic in $\phi$ have dimension $3-\frac32\varepsilon+\text{O}(\varepsilon^2)$, which implies that the deformation \eqref{eq:gendef} is relevant at the trivial ($h_{ijk}=0$) defect. If we deform an interacting bulk by \eqref{eq:gendef}, however, then the $\phi_i\phi_j\phi_k$ operators will organise themselves into irreducible representations of the bulk symmetry at the trivial defect, whose dimensions may be below, at or above $3-\varepsilon$ at order $\varepsilon$. Therefore, such deformations localised on an interface may be relevant, marginal or irrelevant at leading order. Regardless of their relevance in the RG sense, they can all be analysed perturbatively and we will indeed consider below deforming operators that include marginal and irrelevant ones. In essence, we are exploring the space of nontrivial interfaces that lie perturbatively close to the trivial one.

In that space, we typically find interacting interface fixed points with relevant operators cubic in $\phi$. These fixed points are thus RG unstable. For free and $O(N)$ bulks, however, we find that the $F_4$ family \cite{Cvitanovic:2008zz} gives interacting interfaces that are RG stable, realising $SO(3), SU(3), Sp(6)$ and $F_4$ global symmetry with $N=5,8,16,24$ scalar fields, respectively. (To be precise, by RG stable we mean that the symmetry-preserving cubic in $\phi$ deformations localised on the interface are irrelevant. Of course $\phi^2$ localised on the interface is a strongly relevant quadratic operator.) These interfaces are the only interacting RG stable ones we were able to find analytically, but others may be hiding in our numerical results.

The paper is organised as follows. In section \ref{app:MS}, we present in more detail the model and derive the one-loop interface beta functions. In section \ref{sec:analytic}, we compute fixed points analytically for specific defect and bulk symmetries. We also discuss the link between zero eigenvalues of the stability matrix and breaking of the bulk symmetry group. In section \ref{sec:numerics}, we present numerical results for fixed points at low values of $N$ and bulk symmetry $O(N)$, hypercubic, biconical and hypertetrahedral, as well as decoupled and free bulk theories. Finally, in appendix \ref{ap:2loops} we give (incomplete) results on the computation of the two-loop interface beta functions.

\section{Model and beta functions}
\label{app:MS}

We will study a multiscalar model with quartic interactions in a $(d+1)$-dimensional bulk and cubic interactions localised on a $d$-dimensional interface. It is a generalisation of the model studied in \cite{Harribey:2023xyv} and is defined by the action
\begin{equation} \label{eq:action}
		S =  \int d^{\lsp d} \vec x \int dy \, \big(\tfrac{1}{2} \partial^{\mu}\phi_{i} \partial_{\mu}\phi_{i} + \tfrac{1}{4!}\lambda_{ijkl}\lsp\phi_{i} \phi_{j} \phi_{k} \phi_{l}\big)  +  \int d^{\lsp d}\vec x\,
		\tfrac{1}{3!}h_{ijk} \lsp
		\phi_{i} \phi_{j} \phi_{k} \, ,
\end{equation}
where the indices take values from 1 to $N$ and a summation over repeated indices is implicit. The  coupling tensors $\lambda_{ijkl}$ and $h_{ijk}$ are symmetric, thus corresponding in general to $\binom{N+3}{4}$ and $\binom{N+2}{3}$ couplings, respectively. The interface interactions are marginal in dimension three, while the bulk interactions are marginal in dimension four. To allow a perturbative treatment we will thus set $d=3-\varepsilon$ in the following.

One might also consider deformations with operators involving derivatives perpendicular to the interface, namely $\partial_\perp^2\phi_i$ and $\phi_i\partial_\perp\phi_j$. The bulk equation of motion relates the operators $\partial_\perp^2\phi_i$ to $\phi_i\phi_j\phi_k$ and defect descendants. Thus, the operators $\partial_\perp^2\phi_i$ are essentially included in \eqref{eq:action}. The operators $\phi_i\partial_\perp\phi_j$ are independent of the ones appearing in \eqref{eq:action}, but are odd under reflections perpendicular to the interface. Thus, we can  consistently set their couplings to zero, but they remain as non-trivial operators in the interface CFTs we find.

The propagator of the free theory is given by
\begin{align}
\langle\phi_{i}(x_1)\phi_{j}(x_2)\rangle=\delta_{ij}\int\,\frac{d^{\lsp d+1}x}{(2\pi)^{d+1}}\frac{e^{i p\cdot x_{12}}}{p^2} =\delta_{ij}\frac{C_\phi}{|x_{12}|^{d-1}}\,,\qquad C_\phi=\frac{\Gamma\left(\frac{d-1}{2}\right)}{4\pi^{\frac{d+1}{2}}}\,,\qquad x_{12}= x_1- x_2\,.
\end{align}
The free interface-to-bulk propagator $K_{ij}$ and the free interface propagator $G_{ij}$ in momentum space are obtained by Fourier-transforming the parallel but not the perpendicular coordinates to the defect. They were computed in \cite{Harribey:2023xyv} and they are given by
\begin{align}\label{eq:propagators}
K_{ij}(\vec{p}, y) = \delta_{ij}\frac{ e^{-|\vec{p}||y|}}{2|\vec{p}|} \,,\qquad G_{ij}(\vec p)=\delta_{ij}\frac{1}{2|\vec{p}|}\,.
\end{align}

We will compute the beta functions in dimensional regularisation in $d=3-\varepsilon$ using the minimal subtraction scheme.  As is standard, renormalised fields are defined by $\phi_{i,B}=(Z^{1/2})_{ij}\phi_{j,R}$, where the subscripts $B$ and $R$ denote bare and renormalised, respectively. (Such subscripts are omitted below, except where strictly necessary.)

The theory in the bulk will not be modified by the interface interactions. The wavefunction renormalisation factor $Z_{ij}$ is thus the same as for the usual quartic multiscalar model, starting at two loops
\begin{equation}
Z_{ij}=\delta_{ij}-\frac{1}{12\varepsilon}\lambda_{iklm}\lambda_{jklm}+\text{O}(\lambda^3)\,,
\label{eq:wfbulk}
\end{equation}
where we have rescaled the renormalised coupling as $\lambda_{ijkl} \rightarrow 16\pi^{2}\lambda_{ijkl}$. Furthermore, the beta functions for a quartic multiscalar model at one loop are given by the well-known expression
\begin{equation} \label{eq:beta4-general}
		\beta_{ijkl} \, = \, - \, \varepsilon \lambda_{ijkl}
		\, + \, \left(\lambda_{ijmn}\lambda_{mnkl} + 2 \textrm{ perms} \right)  \, ,
\end{equation}
where the ``2 perms'' notation captures the two terms obtained by permuting the free indices in non-equivalent ways.

To compute the beta functions of the interface couplings we require finiteness of a three-point function in the presence of the interface. We find that this is much simpler than determining the beta function by requiring finiteness of the one-point function $\langle\phi_{i}\phi_j\phi_k(x)\rangle$ in the presence of the interface, as would be the natural extension of line and surface defect calculations found in the literature. More specifically, we write the bare defect coupling in terms of the renormalised one as
\begin{equation}
    h_{ijk,B}=\mu^{\varepsilon/2}\left(h_{ijk,R}+\delta h_{ijk}\right) \, ,
\end{equation}
where $\mu$ is an arbitrary mass scale that renders the renormalised coupling $h_{ijk,R}$ dimensionless. The counterterm $\delta h_{ijk}$ has an expansion as a Laurent series in $\varepsilon$,
\begin{equation}\label{eq:defcoupren}
    \delta h_{ijk}=\sum_{n \geq 1} \frac{Z_{ijk}^{(n)}(h,\lambda)}{\varepsilon^n} \, .
\end{equation}
The coefficients $Z^{(n)}_{ijk}(h,\lambda)$ can be obtained by requiring that the three-point function
\begin{equation}
    \Gamma_{ijk}(\vec{p}_1,\vec{p}_2,\vec{p}_3,y) =\delta^{(d)}(\vec{p}_1+\vec{p}_2+\vec{p}_3)\langle\phi_{i}(\vec{p}_1,y)\phi_{j}(\vec{p}_2,y)\phi_k(\vec{p}_3,y)\rangle
\end{equation}
in the presence of the defect is finite. To simplify calculations we take $\vec{p}_3=\vec{0}$ and $\vec{p}_1=-
\vec{p}_2=\vec{p}$. To extract divergences that can be absorbed into a redefinition of the defect coupling as in \eqref{eq:defcoupren}, it suffices to consider the amputated three-point function, whose expected form is
\begin{equation}
    \hat{\Gamma}_{ijk}=h_{ijk,B}+\sum_{\ell\geq1}\frac{1}{|\vec{p}|^{\ell\varepsilon}}\mathcal{A}^{(\ell)}_{ijk}\,,
\label{eq:3pointbare}
\end{equation}
where $\mathcal{A}^{(\ell)}_{ijk}$ is a sum of tensor structures in the bare couplings at loop order $\ell$.

The diagrams contributing to the one-loop three-point function as well as their amplitudes were determined in \cite{Harribey:2023xyv}. Setting $|\vec{p}|=\mu$, the three-point function at one loop is given by
\begin{equation}
    \mu^{-\varepsilon}\mathcal{A}^{(1)}_{ijk}=B\big(\lambda_{ijlm}h_{klm}+ 2 \text{ perms} \big) \, + T\lsp h_{ilm}h_{jln}h_{kmn} \,,
\end{equation}
where
\begin{equation}\label{eq:Bcomp}
    B=\begin{tikzpicture}[baseline=(vert_cent.base), square/.style={regular polygon,regular polygon sides=4}]
        \node at (0,0) [square,draw,fill=white,inner sep=1.2pt,outer sep=0pt]  (l) {};
        \node at (1.2,0) [circle,draw,fill=white,inner sep=1.2pt,outer sep=0pt] (r) {};
        \draw[very thick, densely dashed] (l) to [out=30,in=150] (r);
        \draw[very thick, densely dashed] (l) to [out=-30,in=-150] (r);
        \draw[densely dashed] (l)--++(150:0.4cm);
        \draw[densely dashed] (l)--++(-150:0.4cm);
        \draw[very thick,densely dashed] (r)--++(0:0.4cm);
        \node[inner sep=0pt,outer sep=0pt] (vert_cent) at (0,0) {$\phantom{\cdot}$};
    \end{tikzpicture}
    =-\frac{\mu^{-\varepsilon}}{16\pi^2\varepsilon} +\text{O}(\varepsilon^0) \, ,
\end{equation}
and
\begin{equation}
T=\begin{tikzpicture}[baseline=(vert_cent.base)]
        \node at (90:0.6cm) [circle,draw,fill=white,inner sep=1.2pt,outer sep=0pt]  (t) {};
        \node at (210:0.6cm) [circle,draw,fill=white,inner sep=1.2pt,,outer sep=0pt] (l) {};
        \node at (330:0.6cm) [circle,draw,fill=white,inner sep=1.2pt,,outer sep=0pt] (r) {};
        \draw[very thick] (t)--(l);
        \draw[very thick] (l)--(r);
        \draw[very thick] (r)--(t);
        \node (vert_cent) at (current bounding box.center) {};
        \draw[very thick,densely dashed] (t)--++(90:0.4cm);
        \draw[very thick,densely dashed] (l)--++(210:0.4cm);
        \draw[very thick,densely dashed] (r)--++(330:0.4cm);
    \end{tikzpicture}
=\frac{\mu^{-\varepsilon}}{16\pi^{2}\varepsilon} +\text{O}(\varepsilon^0)\,.
\end{equation}
In our graphical representation, thin dashed propagators are in the bulk, thick dashed propagators are from the bulk to the defect ($K$ in \eqref{eq:propagators}), and solid ones are purely on the defect ($G$ in \eqref{eq:propagators}). Additionally, square vertices are in the bulk, while circular ones are on the defect. The counterterm $Z^{(1)}_{ijk}$ is then simply given by
\begin{equation}\label{eq:Z1oneloop}
    Z^{(1)}_{ijk}= \frac{1}{16\pi^2\varepsilon}\big(\lambda_{ijlm}h_{klm}+ 2 \text{ perms} \big) \, - \frac{1}{16\pi^{2}\varepsilon}h_{ilm}h_{jln}h_{kmn}  \, .
\end{equation}
In minimal subtraction the beta function coefficients can be read off from these simple-pole residues \cite{tHooft:1973mfk}. In particular, in our case the beta functions are given by
\begin{equation}
    \beta_{ijk}=-\frac{\varepsilon}{2}h_{ijk}+\frac{1}{2}\left( h_{lmn}\frac{\partial}{\partial h_{lmn}}+2\lambda_{lmnp}\frac{\partial}{\partial \lambda_{lmnp}}-1 \right)Z^{(1)}_{ijk} \, .
    \label{eq:genericbetaMS}
\end{equation}
With our result \eqref{eq:Z1oneloop} we finally obtain the one-loop interface beta function
\begin{align}\label{eq:betaone}
\beta_{ijk} &= -\frac{\varepsilon}{2} h_{ijk}
 +  \big(\lambda_{ijlm}h_{klm}+ 2 \textrm{ perms} \big) -\frac{1}{4}h_{ilm}h_{jln}h_{kmn} \, ,
\end{align}
where we have rescaled the interface coupling as $h_{ijk} \rightarrow 2\pi h_{ijk}$ (and the bulk coupling as $\lambda_{ijkl} \rightarrow 16\pi^{2}\lambda_{ijkl}$ as above). At this order, the beta function is gradient, i.e.
\begin{equation}
    \beta_{ijk}=\frac{\partial H}{\partial h_{ijk}}\,,\qquad H=-\frac14\varepsilon \lsp h_{ijk}h_{ijk}+\frac32\lambda_{ijkl}h_{ijm}h_{klm} - \frac{1}{16} h_{ijk}h_{ilm}h_{jln}h_{kmn}\,.
\end{equation}

\begin{figure}[H]
\centering
\captionsetup[subfigure]{labelformat=empty}
\subfloat[$I_1$]{\begin{tikzpicture}[baseline=(vert_cent.base),scale=0.4]
\node at (4,0) [circle,draw,fill=white,inner sep=1.2pt,outer sep=0pt]  (a) {};
\node at (2,3.46) [circle,draw,fill=white,inner sep=1.2pt,outer sep=0pt]  (b) {};
\node at (0,0) [circle,draw,fill=white,inner sep=1.2pt,outer sep=0pt]  (c) {};
\node at (3,1.73) [circle,draw,fill=white,inner sep=1.2pt,outer sep=0pt]  (d) {};
\node at (1,1.73) [circle,draw,fill=white,inner sep=1.2pt,outer sep=0pt]  (e) {};
\node (vert_cent) at (current bounding box.center) {};
\draw [very thick] (a)--(d) ;
\draw [very thick] (d)--(b) ;
\draw [very thick] (c)--(e) ;
\draw [very thick] (e)--(b);
\draw [very thick] (e)--(d);
\draw [very thick] (c)--(a);
\draw[very thick,densely dashed] (b)--++(90:1cm);
\draw[very thick,densely dashed] (c)--++(210:1cm);
\draw[very thick,densely dashed] (a)--++(330:1cm);
\end{tikzpicture}} \hspace{1cm}
\subfloat[$I_2$]{\begin{tikzpicture}[scale=0.4]
\node at (4,0) [circle,draw,fill=white,inner sep=1.2pt,outer sep=0pt]  (a) {};
\node at (2,3.46) [circle,draw,fill=white,inner sep=1.2pt,outer sep=0pt]  (b) {};
\node at (0,0) [circle,draw,fill=white,inner sep=1.2pt,outer sep=0pt]  (c) {};
\node at (3.37,1.08) [circle,draw,fill=white,inner sep=1.2pt,outer sep=0pt]  (d) {};
\node at (2.62,2.38) [circle,draw,fill=white,inner sep=1.2pt,outer sep=0pt]  (e) {};
\draw [very thick] (a)-- (d) ;
\draw [very thick] (e)-- (b) ;
\draw [very thick] (d) to [bend right] (e) ;
\draw [very thick] (d) to [bend left] (e) ;
\draw [very thick] (c)--(b) ;
\draw [very thick] (c)--(a);
\draw[very thick,densely dashed] (b)--++(90:1cm);
\draw[very thick,densely dashed] (c)--++(210:1cm);
\draw[very thick,densely dashed] (a)--++(330:1cm);
\end{tikzpicture}} \hspace{1cm}
\subfloat[$I_3$]{\begin{tikzpicture}[baseline=(vert_cent.base),scale=0.4]
\node at (4,0) [circle,draw,fill=white,inner sep=1.2pt,outer sep=0pt]  (a) {};
\node at (2,3.46) [circle,draw,fill=white,inner sep=1.2pt,outer sep=0pt]  (b) {};
\node at (0,0) [circle,draw,fill=white,inner sep=1.2pt,outer sep=0pt]  (c) {};
\node at (3,1.73) [circle,draw,fill=white,inner sep=1.2pt,outer sep=0pt]  (d) {};
\node at (1,1.73) [circle,draw,fill=white,inner sep=1.2pt,outer sep=0pt]  (e) {};
\draw [very thick] (e)-- (a);
\draw [white, line width=0.11cm] (c)-- (d);
\draw [very thick] (a)-- (d) ;
\draw [very thick] (d)-- (b) ;
\draw [very thick] (c)-- (e) ;
\draw [very thick] (e)-- (b);
\draw [very thick] (c)-- (d);
\draw[very thick, densely dashed] (b)--++(90:1cm);
\draw[very thick, densely dashed] (c)--++(210:1cm);
\draw[very thick, densely dashed] (a)--++(330:1cm);
\end{tikzpicture}}
\caption{Two-loop graphs with only cubic couplings}
\label{fig:2loopcubic}
\end{figure}
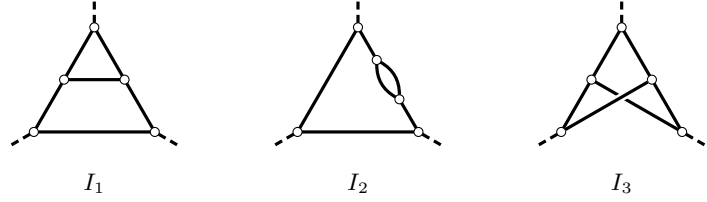

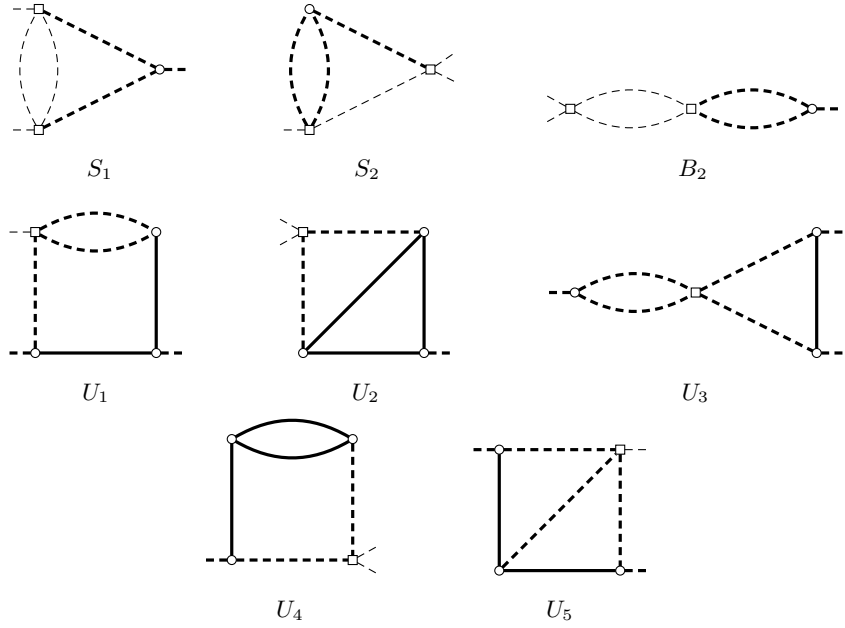
\begin{figure}[H]
\centering
\captionsetup[subfigure]{labelformat=empty}
\subfloat[$S_1$]{\begin{tikzpicture}[baseline=(vert_cent.base),square/.style={regular polygon,regular polygon sides=4},scale=0.4]
\node at (0,4) [square,draw,fill=white,inner sep=1.2pt,outer sep=0pt]  (a) {};
\node at (0,0) [square,draw,fill=white,inner sep=1.2pt,outer sep=0pt]  (b) {};
\node at (4,2) [circle,draw,fill=white,inner sep=1.2pt,outer sep=0pt]  (c) {};
\draw [densely dashed] (b) to [bend right] (a) ;
\draw [densely dashed] (b) to [bend left] (a) ;
\draw[very thick,densely dashed] (b)--(c);
\draw[very thick,densely dashed] (a)--(c);
\draw[densely dashed] (b)--(-1,0);
\draw[densely dashed] (a)--(-1,4);
\draw[very thick, densely dashed] (c)--(5,2);
\end{tikzpicture}} \hspace{1cm}
\subfloat[$S_2$]{\begin{tikzpicture}[baseline=(vert_cent.base),square/.style={regular polygon,regular polygon sides=4},scale=0.4]
\node at (0,4) [circle,draw,fill=white,inner sep=1.2pt,outer sep=0pt]  (a) {};
\node at (0,0) [square,draw,fill=white,inner sep=1.2pt,outer sep=0pt]  (b) {};
\node at (4,2) [square,draw,fill=white,inner sep=1.2pt,outer sep=0pt]  (c) {};
\node (vert_cent) at (current bounding box.center) {};
\draw [very thick,densely dashed] (b) to [bend right] (a) ;
\draw [very thick,densely dashed] (b) to [bend left] (a) ;
\draw[densely dashed] (b)--(c);
\draw[very thick,densely dashed] (a)--(c);
\draw[densely dashed] (b)--(-1,0);
\draw[densely dashed] (c)--(4.8,1.6);
\draw[densely dashed] (c)--(4.8,2.55);
\end{tikzpicture}} \hspace{1cm}
\subfloat[$B_2$]{\begin{tikzpicture}[baseline=(vert_cent.base),square/.style={regular polygon,regular polygon sides=4},scale=0.4]
\node at (0,0) [square,draw,fill=white,inner sep=1.2pt,outer sep=0pt]  (a) {};
\node at (4,0) [square,draw,fill=white,inner sep=1.2pt,outer sep=0pt]  (b) {};
\node at (8,0) [circle,draw,fill=white,inner sep=1.2pt,outer sep=0pt]  (c) {};
\draw [densely dashed] (a) to [bend right] (b) ;
\draw [densely dashed] (a) to [bend left] (b) ;
\draw [very thick,densely dashed] (c) to [bend right] (b) ;
\draw [very thick,densely dashed] (c) to [bend left] (b) ;
\draw[densely dashed] (a)--(-0.9,-0.5);
\draw[densely dashed] (a)--(-0.9,0.5);
\draw[very thick,densely dashed] (c)--(9,0);
\end{tikzpicture}} \\[6pt]
\subfloat[$U_1$]{\begin{tikzpicture}[baseline=(vert_cent.base),square/.style={regular polygon,regular polygon sides=4},scale=0.4]
\node at (0,0) [circle,draw,fill=white,inner sep=1.2pt,outer sep=0pt]  (a) {};
\node at (0,4) [square,draw,fill=white,inner sep=1.2pt,outer sep=0pt]  (b) {};
\node at (4,4) [circle,draw,fill=white,inner sep=1.2pt,outer sep=0pt]  (c) {};
\node at (4,0) [circle,draw,fill=white,inner sep=1.2pt,outer sep=0pt]  (d) {};
\draw [very thick,densely dashed] (b) to [bend right] (c) ;
\draw [very thick,densely dashed] (b) to [bend left] (c) ;
\draw[very thick,densely dashed] (a)--(b);
\draw[very thick] (a)--(d);
\draw[very thick] (c)--(d);
\draw[very thick,densely dashed] (a)--(-1,0);
\draw[densely dashed] (b)--(-1,4);
\draw[very thick,densely dashed] (d)--(5,0);
\end{tikzpicture}} \hspace{1cm}
\subfloat[$U_2$]{\begin{tikzpicture}[baseline=(vert_cent.base),square/.style={regular polygon,regular polygon sides=4},scale=0.4]
\node at (0,0) [circle,draw,fill=white,inner sep=1.2pt,outer sep=0pt]  (a) {};
\node at (0,4) [square,draw,fill=white,inner sep=1.2pt,outer sep=0pt]  (b) {};
\node at (4,4) [circle,draw,fill=white,inner sep=1.2pt,outer sep=0pt]  (c) {};
\node at (4,0) [circle,draw,fill=white,inner sep=1.2pt,outer sep=0pt]  (d) {};
\node (vert_cent) at (current bounding box.center) {};
\draw[very thick,densely dashed] (a)--(b);
\draw[very thick] (a)--(d);
\draw[very thick] (c)--(d);
\draw[very thick,densely dashed] (b)--(c);
\draw[very thick] (a)--(c);
\draw[densely dashed] (b)--(-0.9,3.5);
\draw[densely dashed] (b)--(-0.9,4.5);
\draw[very thick,densely dashed] (d)--(5,0);
\end{tikzpicture}}
\hspace{1cm}
\subfloat[$U_3$]{\begin{tikzpicture}[baseline=(vert_cent.base),square/.style={regular polygon,regular polygon sides=4},scale=0.4]
\node at (0,0) [circle,draw,fill=white,inner sep=1.2pt,outer sep=0pt]  (a) {};
\node at (4,0) [square,draw,fill=white,inner sep=1.2pt,outer sep=0pt]  (b) {};
\node at (8,2) [circle,draw,fill=white,inner sep=1.2pt,outer sep=0pt]  (c) {};
\node at (8,-2) [circle,draw,fill=white,inner sep=1.2pt,outer sep=0pt]  (d) {};
\node (vert_cent) at (current bounding box.center) {};
\draw [very thick,densely dashed] (a) to [bend right] (b) ;
\draw [very thick,densely dashed] (a) to [bend left] (b) ;
\draw[very thick,densely dashed] (b)--(c);
\draw[very thick,densely dashed] (b)--(d);
\draw[very thick] (c)--(d);
\draw[very thick,densely dashed] (a)--(-1,0) ;
\draw[very thick,densely dashed] (c)--(9,2) ;
\draw[very thick,densely dashed] (d)--(9,-2) ;
\end{tikzpicture}}  \\
\subfloat[$U_4$]{\begin{tikzpicture}[baseline=(vert_cent.base),square/.style={regular polygon,regular polygon sides=4},scale=0.4]
\node at (0,0) [circle,draw,fill=white,inner sep=1.2pt,outer sep=0pt]  (a) {};
\node at (0,4) [circle,draw,fill=white,inner sep=1.2pt,outer sep=0pt]  (b) {};
\node at (4,4) [circle,draw,fill=white,inner sep=1.2pt,outer sep=0pt]  (c) {};
\node at (4,0) [square,draw,fill=white,inner sep=1.2pt,outer sep=0pt]  (d) {};
\node (vert_cent) at (current bounding box.center) {};
\draw [very thick] (b) to [bend right] (c) ;
\draw [very thick] (b) to [bend left] (c) ;
\draw[very thick] (a)--(b);
\draw[very thick,densely dashed] (a)--(d);
\draw[very thick,densely dashed] (c)--(d);
\draw[very thick,densely dashed] (a)--(-1,0);
\draw[densely dashed] (d)--(4.9,-0.5);
\draw[densely dashed] (d)--(4.9,0.5);
\end{tikzpicture}} \hspace{1cm}
\subfloat[$U_5$]{\begin{tikzpicture}[baseline=(vert_cent.base),square/.style={regular polygon,regular polygon sides=4},scale=0.4]
\node at (0,0) [circle,draw,fill=white,inner sep=1.2pt,outer sep=0pt]  (a) {};
\node at (0,4) [circle,draw,fill=white,inner sep=1.2pt,outer sep=0pt]  (b) {};
\node at (4,4) [square,draw,fill=white,inner sep=1.2pt,outer sep=0pt]  (c) {};
\node at (4,0) [circle,draw,fill=white,inner sep=1.2pt,outer sep=0pt]  (d) {};
\node (vert_cent) at (current bounding box.center) {};
\draw[very thick] (a)--(b);
\draw[very thick] (a)--(d);
\draw[very thick,densely dashed] (c)--(d);
\draw[very thick,densely dashed] (b)--(c);
\draw[very thick,densely dashed] (a)--(c);
\draw[very thick,densely dashed] (b)--(-1,4);
\draw[densely dashed] (c)--(5,4);
\draw[very thick,densely dashed] (d)--(5,0);
\end{tikzpicture}}
\caption{Two-loop graphs with both cubic and quartic couplings.}
\label{fig:2loopmixed}
\end{figure}

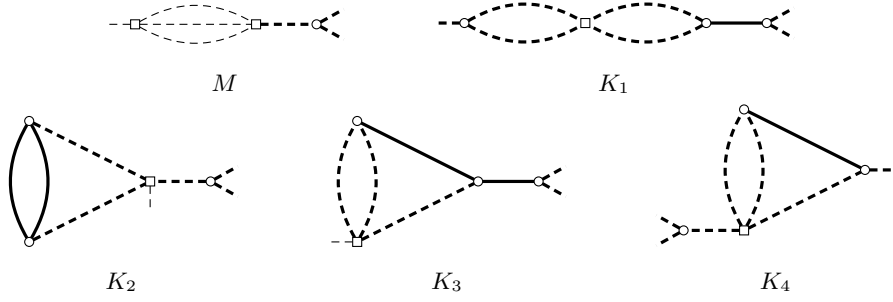
\begin{figure}[H]
\centering
\captionsetup[subfigure]{labelformat=empty}
\subfloat[$M$]{\begin{tikzpicture}[baseline=(vert_cent.base),square/.style={regular polygon,regular polygon sides=4},scale=0.4]
\node at (0,0) [square,draw,fill=white,inner sep=1.2pt,outer sep=0pt]  (a) {};
\node at (4,0) [square,draw,fill=white,inner sep=1.2pt,outer sep=0pt]  (b) {};
\node at (6,0) [circle,draw,fill=white,inner sep=1.2pt,outer sep=0pt]  (c) {};
\node (vert_cent) at (current bounding box.center) {};
\draw [densely dashed] (a) to [bend right] (b) ;
\draw [densely dashed] (a) to [bend left] (b) ;
\draw [densely dashed] (a) -- (b) ;
\draw [very thick,densely dashed] (b) -- (c) ;
\draw[densely dashed] (a)--(-1,0);
\draw[very thick,densely dashed] (c)--(6.9,0.5);
\draw[very thick,densely dashed] (c)--(6.9,-0.5);
\end{tikzpicture}} \hspace{1cm}
\subfloat[$K_1$]{\begin{tikzpicture}[baseline=(vert_cent.base),square/.style={regular polygon,regular polygon sides=4},scale=0.4]
\node at (0,0) [circle,draw,fill=white,inner sep=1.2pt,outer sep=0pt]  (a) {};
\node at (4,0) [square,draw,fill=white,inner sep=1.2pt,outer sep=0pt]  (b) {};
\node at (8,0) [circle,draw,fill=white,inner sep=1.2pt,outer sep=0pt]  (c) {};
\node at (10,0) [circle,draw,fill=white,inner sep=1.2pt,outer sep=0pt]  (d) {};
\node (vert_cent) at (current bounding box.center) {};
\draw [very thick,densely dashed] (a) to [bend right] (b) ;
\draw [very thick,densely dashed] (a) to [bend left] (b) ;
\draw [very thick,densely dashed] (c) to [bend right] (b) ;
\draw [very thick,densely dashed] (c) to [bend left] (b) ;
\draw[very thick] (c)--(d);
\draw[very thick,densely dashed] (a)--(-1,0);
\draw[very thick,densely dashed] (d)--(10.9,0.5);
\draw[very thick,densely dashed] (d)--(10.9,-0.5);
\end{tikzpicture}} \\
\subfloat[$K_2$]{\begin{tikzpicture}[baseline=(vert_cent.base),square/.style={regular polygon,regular polygon sides=4},scale=0.4]
\node at (0,0) [circle,draw,fill=white,inner sep=1.2pt,outer sep=0pt]  (a) {};
\node at (0,4) [circle,draw,fill=white,inner sep=1.2pt,outer sep=0pt]  (b) {};
\node at (4,2) [square,draw,fill=white,inner sep=1.2pt,outer sep=0pt]  (c) {};
\node at (6,2) [circle,draw,fill=white,inner sep=1.2pt,outer sep=0pt]  (d) {};
\node (vert_cent) at (current bounding box.center) {};
\draw [very thick] (a) to [bend right] (b) ;
\draw [very thick] (a) to [bend left] (b) ;
\draw[very thick,densely dashed] (a)--(c);
\draw[very thick,densely dashed] (b)--(c);
\draw[very thick,densely dashed] (d)--(c);
\draw[densely dashed] (c)--(4,1);
\draw[very thick,densely dashed] (d)--(6.9,2.5);
\draw[very thick,densely dashed] (d)--(6.9,1.5);
\end{tikzpicture}} \hspace{1cm}
\subfloat[$K_3$]{\begin{tikzpicture}[baseline=(vert_cent.base),square/.style={regular polygon,regular polygon sides=4},scale=0.4]
\node at (0,0) [square,draw,fill=white,inner sep=1.2pt,outer sep=0pt]  (a) {};
\node at (0,4) [circle,draw,fill=white,inner sep=1.2pt,outer sep=0pt]  (b) {};
\node at (4,2) [circle,draw,fill=white,inner sep=1.2pt,outer sep=0pt]  (c) {};
\node at (6,2) [circle,draw,fill=white,inner sep=1.2pt,outer sep=0pt]  (d) {};
\node (vert_cent) at (current bounding box.center) {};
\draw [very thick,densely dashed] (a) to [bend right] (b) ;
\draw [very thick,densely dashed] (a) to [bend left] (b) ;
\draw[very thick,densely dashed] (a)--(c);
\draw[very thick] (b)--(c);
\draw[very thick] (d)--(c);
\draw[densely dashed] (a)--(-1,0);
\draw[very thick,densely dashed] (d)--(6.9,2.5);
\draw[very thick,densely dashed] (d)--(6.9,1.5);
\end{tikzpicture}} \hspace{1cm}
\subfloat[$K_4$]{\begin{tikzpicture}[baseline=(vert_cent.base),square/.style={regular polygon,regular polygon sides=4},scale=0.4]
\node at (0,0) [square,draw,fill=white,inner sep=1.2pt,outer sep=0pt]  (a) {};
\node at (0,4) [circle,draw,fill=white,inner sep=1.2pt,outer sep=0pt]  (b) {};
\node at (4,2) [circle,draw,fill=white,inner sep=1.2pt,outer sep=0pt]  (c) {};
\node at (-2,0) [circle,draw,fill=white,inner sep=1.2pt,outer sep=0pt]  (d) {};
\node (vert_cent) at (current bounding box.center) {};
\draw [very thick,densely dashed] (a) to [bend right] (b) ;
\draw [very thick,densely dashed] (a) to [bend left] (b) ;
\draw[very thick,densely dashed] (a)--(c);
\draw[very thick] (b)--(c);
\draw[very thick,densely dashed] (a)--(d);
\draw[very thick,densely dashed] (c)--(5,2);
\draw[very thick,densely dashed] (d)--(-2.9,0.5);
\draw[very thick,densely dashed] (d)--(-2.9,-0.5);
\end{tikzpicture}}
\caption{One-particle reducible graphs contributing to the interface three-point function}
\label{fig:2pt}
\end{figure}

At two loops there is a variety of diagrams that contribute to the beta function. Using our graphical notation from above, these are given in Figs.\ \ref{fig:2loopcubic}, \ref{fig:2loopmixed} and \ref{fig:2pt}. While we have not been able to compute all of their contributions to the interface beta function, we summarise the results we have obtained in appendix~\ref{ap:2loops}.

Here we would like to mention a trick that allows the calculation of contributions to the beta function linear in the interface coupling $h$. At the trivial defect, $h=0$, eigenvalues of the stability matrix $\partial\beta/\partial h$ are directly related to the anomalous dimensions of the nearly marginal deforming operators. If the bulk theory is also free, this observation is of no use, but if it is not it sets constraints on the coefficients of the terms linear in $h$ that appear in the defect beta function.

For example, if the bulk CFT is the $O(N)$ model and we work at one loop, then there are two linearly-independent beta functions \cite{Harribey:2023xyv} and, neglecting terms cubic in the couplings $h_1$, $h_2$, which are irrelevant for the argument we are making here, we have
\begin{equation}
    \beta_{1}\supset\Big(\frac{2\alpha-1}{2}h_1+\frac{\alpha}{N+8}h_2\Big)\varepsilon\,,\qquad \beta_2\supset-\frac{(N+8-12\lsp\alpha)}{2(N+8)}\lsp h_2\lsp \varepsilon\,,
\end{equation}
where $\alpha$ is the beta function coefficient that follows from \eqref{eq:Bcomp}; $\alpha=1$ from \eqref{eq:betaone}. The matrix $\partial\beta_a/\partial h_b$, $a,b=1,2$, evaluated at the trivial defect, $h_1=h_2=0$, is upper-triangular and thus has the obvious eigenvalues
\begin{equation}
    \kappa_1=\Big(\alpha-\frac12\Big)\varepsilon\,,\qquad
    \kappa_2=-\frac{N+8-12\lsp\alpha}{N+8}\lsp\varepsilon\,.
\end{equation}
These eigenvalues then determine the scaling dimensions of bulk operators one uses to deform away from the trivial defect. In our case these are cubic operators in particular $O(N)$ irreps, namely vector and three-index traceless symmetric, and their dimensions are known from bulk computations to be\footnote{$\Delta_1=4-\varepsilon-\Delta_\phi$ corresponds to the equation of motion operator and $\Delta_2$ to the three-index traceless symmetric of the $O(N)$ model.}
\begin{equation}\label{eq:defops}
    \Delta_1=3-\frac12\lsp\varepsilon+\text{O}(\varepsilon^2)\,,\qquad \Delta_2=3-\frac32\lsp\varepsilon+\frac{6}{N+8}\lsp\varepsilon+\text{O}(\varepsilon^2)\,,
\end{equation}
respectively~\cite{Henriksson:2022rnm}. Neglecting terms of order $\varepsilon^2$ and higher, it must be that
\begin{equation}
    \kappa_{1,2}=\Delta_{1,2}-(3-\varepsilon)\,,
\end{equation}
which implies that $\alpha=1$ consistently with the explicit computation of the loop diagram \eqref{eq:Bcomp}.

At two loops there are four contributions to the beta function linear in $h_{1,2}$. Using the order $\varepsilon^2$ contributions to \eqref{eq:defops} given in \cite{Henriksson:2022rnm} allows us to determine all four corresponding coefficients. These are otherwise given by rather complicated integrals, three of which we have not been able to evaluate analytically.

\section{Analytic results}
\label{sec:analytic}
\subsection{Free bulk}
Fixed points of \eqref{eq:betaone} are found by first fixing $\lambda_{ijkl}$, which corresponds to a determination of the bulk universality class. As a first example, let us choose the bulk to be free. Then, interface fixed points correspond to solutions of
\begin{equation}\label{eq:betafree}
  \beta_{ijk}=-\tfrac12\varepsilon\lsp h_{ijk}
  -\tfrac{1}{4}h_{ilm}h_{jln}h_{kmn}=0\,.
\end{equation}
Fixed points of this equation are not easy to analyse in full generality, so we discuss here some examples that are determined by the symmetry we would like to preserve on the interface. The bulk free theory has $O(N)$ symmetry and we may choose to preserve different subgroups of that.

The case of $O(N-1)$ on the interface was extensively analysed in~\cite{Harribey:2023xyv}. It was found that for $N\leq 2$ there are no unitary fixed points, while for $N>2$ there is one pair of physically equivalent unitary fixed points for each $N$. These fixed points have relevant operators, so they are unstable.

An interesting class of examples with one defect coupling arises when $h_{ijk}=\mathdutchcal{h}\lsp d_{ijk}$, with $d_{ijk}$ a symmetric traceless tensor. If $d_{ijk}$ is the only such tensor and there are no primitive two-index invariant tensors beyond $\delta_{ij}$, then we may impose
\begin{equation}
    d_{ikl}d_{jkl}=r\lsp\delta_{ij}\,,\quad r>0\,,\qquad d_{ilm}d_{jln}d_{kmn}=s\lsp d_{ijk}\,,
\end{equation}
and \eqref{eq:betafree} becomes
\begin{equation}
    \beta_{\mathdutchcal{h}}=-\tfrac12\varepsilon\lsp\mathdutchcal{h}-\tfrac14 s\lsp \mathdutchcal{h}^3=0\,,
\end{equation}
which is solved for $\mathdutchcal{h}=0$ or
\begin{equation}\label{eq:intfreebulk}
    \mathdutchcal{h}^2=-\frac{2}{s}\lsp\varepsilon\,.
\end{equation}
Unitarity requires $s<0$. The constants $r,s$ obey the bound \cite{Osborn:2017ucf}
\begin{equation}\label{eq:boundalbe}
    -\frac{N-2}{2(N+2)}\lsp r\leq s \leq \frac{N-2}{N-1}\lsp r\,.
\end{equation}
At the non-trivial interface fixed point given by \eqref{eq:intfreebulk} the operator $d_{ijk}\phi_i\phi_j\phi_k$ has dimension $3+\text{O}(\varepsilon^2)$ and thus gives rise to an irrelevant perturbation when localised on the interface; this fixed point is RG stable. (At the trivial interface, $d_{ijk}\phi_i\phi_j\phi_k$ has dimension $3-\frac32\varepsilon+\text{O}(\varepsilon^2)$ and corresponds to a relevant perturbation when localised on the interface.)

$S_{N+1}$ symmetry is an example of this construction. To realise it, we introduce $N+1$ vectors $e_i^\alpha$, $\alpha=1,\ldots,N+1$,\footnote{These vectors point to the $N+1$ vertices of an $N$-dimensional hypertetrahedron.} satisfying
\begin{equation}
    \sum_{\alpha} e_i^\alpha=0\,,\qquad \sum_{\alpha}e^\alpha_i e_j^\alpha=\delta_{ij}\,,\qquad e_i^\alpha e_i^\beta=\delta^{\alpha\beta}-\frac{1}{N+1}\,.
    \label{eq:evectors}
\end{equation}
Then,
\begin{equation}
    d_{ijk}=\sum_{\alpha}e_i^\alpha e_j^\alpha e_k^\alpha
\end{equation}
and one may compute \cite[Appendix C]{Osborn:2017ucf}
\begin{equation}\label{eq:snrs}
    r=\frac{N-1}{N+1}\,,\qquad s=\frac{N-2}{N+1}\,.
\end{equation}
The $S_{N+1}$ theory saturates the upper bound of \eqref{eq:boundalbe}. Unitarity requires $N<2$ and so unitary $S_{N+1}$ symmetric fixed points do not exist for a free bulk.

Another example is given by $SU(n)$ symmetry, $n>2$, $N=n^2-1,$ for which $d_{ijk}$ is the standard symmetric tensor and then
\begin{equation}
    r=\frac{n^2-4}{n}\,,\qquad s=\frac{n^2-12}{2n}\,.
\end{equation}
Unitary fixed points can be found for $n^2<12$. In particular, $n=3$, for which $N=8$, gives an $SU(3)$ symmetric interacting interface CFT which, as discussed above, is RG stable.

Beyond the $SU(3)$ example, there are three further examples from the $F_4$ family considered in \cite{Cvitanovic:2008zz},\footnote{Discussions of this family in different settings than considered here can be found in~\cite{Pang:2016xno, Rong:2019qer, Liendo:2021wpo, Jack:2023zjt}.} for which $N=5$, $N=14$ and $N=26$, with symmetry $SO(3)$, $Sp(6)$ and $F_4$, respectively. All of these, like the $SU(3)$ case, saturate the lower bound of \eqref{eq:boundalbe} and thus satisfy
\begin{equation}
    s=-\frac{N-2}{2(N+2)}\lsp r
\end{equation}
for their respective $N$. Since $r>0$, they define unitary interacting CFTs that are RG stable.

We have also sought fixed points with $S_N$ symmetry, for which there are three independent couplings. For this case we do not have a general closed-form solution for the fixed points valid for any $N$. However, all unitary fixed points we have found by demanding $S_N$ symmetry have enhanced $O(N-1)$ symmetry.

\subsection{\texorpdfstring{$O(N)$}{O(N)} bulk}\label{sec:ONbulk}
When the bulk universality class is fixed to the $O(N)$ model, then again \cite{Harribey:2023xyv} found unitary interface fixed points with $O(N-1)$ symmetry for $2\leq N\leq 7$ which were RG unstable, while for $N$ outside this range there were only non-unitary interface fixed points with $O(N-1)$ symmetry.

Considering fixed points with one rank-three symmetric traceless invariant tensor as in the previous section now gives fixed points as solutions of
\begin{equation}
    \beta_{\mathdutchcal{h}}=-\frac{N-4}{2(N+8)}\lsp\varepsilon\lsp\mathdutchcal{h}-\frac14 s\lsp\mathdutchcal{h}^3=0\,.
\end{equation}
If $\mathdutchcal{h}\neq0$ this is solved for
\begin{equation}
    \mathdutchcal{h}^2=-\frac{2(N-4)}{N+8}\frac{1}{s}\lsp\varepsilon\,,
\end{equation}
and at this fixed point the operator $d_{ijk}\phi_i\phi_j\phi_k$ has dimension $3-\frac{12}{N+8}\varepsilon+\text{O}(\varepsilon^2)$. RG stability requires $N>4$ and unitarity in that case imposes $s<0$. (At the trivial interface, $d_{ijk}\phi_i\phi_j\phi_k$ has dimension $3-\frac{3(N+4)}{2(N+8)}\lsp\varepsilon+\text{O}(\varepsilon^2)$ and corresponds to a relevant perturbation when localised on the interface if $N>4$.)

Due to \eqref{eq:snrs}, $S_{N+1}$ symmetry on the interface can be realised on a unitary fixed point only for $N=3$ and this fixed point is RG unstable. Unitary RG stable fixed points exist for $SO(3)$, $SU(3)$, $Sp(6)$ and $F_4$ symmetries, with $N=5,8,14,26$, respectively.

Requiring $S_N$ symmetry on the interface we again find a unitary fixed point only for $N=3$. As it turns out (see Table \ref{tab:O3} below), for $N=3$ the fixed points we have found, namely those with symmetry $O(2)$, $S_4$ and $S_3$, along with two fixed points with symmetry $\mathbb{Z}_2^2$, constitute the full set of unitary interface fixed points that can be obtained for an $O(3)$ bulk.

Let us remark here that for $N=4$ the three-index traceless symmetric operator localised on an interface leads to a marginal perturbation at leading order in $\varepsilon$, as can be seen from $\Delta_2$ in \eqref{eq:defops}. Therefore, one may find a manifold of fixed points at this order. Indeed, this is the case, as there exist solutions of $\beta_{ijk}=0$ for the beta function given in \eqref{eq:betaone} for which the first two terms cancel and the term cubic in $h$ can be set to zero without fixing all couplings. There are then genuine ``conformal manifolds'' at leading order, but we have checked using our two-loop results that they do not survive at next-to-leading order; for this it is enough that the contribution of the $I_3$ diagram in Fig.\ \ref{fig:2loopcubic} to the two-loop interface beta function is non-zero.

\subsection{Hypertetrahedral bulk}
One can additionally consider the bulk to lie at one of the two hypertetrahedral fixed points~\cite{Osborn:2017ucf}, with symmetry $S_{N+1}\times\mathbb{Z}_2$. The bulk interaction tensor is most naturally constructed from the $N+1$ vectors $e_i^\alpha$ satisfying \eqref{eq:evectors}, taking the explicit form
\begin{equation}
    \lambda_{ijkl}=\varepsilon\lambda_{\pm}\left(\delta_{ij}\delta_{kl}+\text{perms}\right)+\varepsilon g_{\pm}\sum_{\alpha}e^\alpha_i e^\alpha_j e^\alpha_k e^\alpha_l\,,
\end{equation}
where
\begin{equation}
\begin{split}
    \lambda_+&=\frac{1}{3(N^2-5N+8)}\,, \qquad g_+=\frac{(N-4)(N+1)}{3(N^2-5N+8)}\,, \\
    \lambda_-&=\frac{1}{3(N+3)}\,, \qquad\qquad\quad g_-=\frac{N+1}{3(N+3)}\,.
\end{split}
\end{equation}

Considering $S_{N+1}$ fixed points with the rank-three symmetric traceless tensor given as before by
\begin{equation}
    h_{ijk}=\mathdutchcal{h}\sum_\alpha e^\alpha_i e^\alpha_j e^\alpha_k\,,
\end{equation}
one finds that fixed points solve
\begin{equation}
    \beta_{\mathdutchcal{h}}=-\frac{\varepsilon}{2}\mathdutchcal{h}+6\varepsilon\lambda_\pm \mathdutchcal{h}+\frac{3(N-1)\varepsilon}{N+1}g \mathdutchcal{h}-\frac{N-2}{4(N+1)}\mathdutchcal{h}^3=0\,.
\end{equation}
This has the non-trivial fixed point
\begin{equation}
    \mathdutchcal{h}^2=\frac{4(N-2)}{N+1}\left(-\frac{1}{2}+6\lambda_\pm+\frac{3(N-1)}{N+1}g_\pm\right)\varepsilon\,,
\end{equation}
which for the two different fixed points becomes
\begin{equation}
    \mathdutchcal{h}_+^2=\frac{2(N-4)(N-1)(N+1)}{(N-2)(N^2-5N+8)}\lsp\varepsilon
\end{equation}
and
\begin{equation}
    \mathdutchcal{h}_-^2=\frac{2(N-1)(N+1)}{N^2+N-6}\lsp\varepsilon\,,
\end{equation}
respectively. One sees that $\mathdutchcal{h}^2_+$ will correspond to a unitary fixed point for $N\geq 4$, while $\mathdutchcal{h}^2_-$ will be unitary for $N>2$. Both fixed points, however, will be RG unstable in these ranges of unitarity, with the cubic operator $d_{ijk}\phi_i\phi_j\phi_k$ having dimension $3-\frac{4}{N^2-5N+8}\varepsilon$ and $3-\frac{4}{N+3}\varepsilon$ at the non-trivial fixed points at $\mathdutchcal{h}_+^2$ and $\mathdutchcal{h}_-^2$, respectively.

We must remark here that for $N=5$ there exists a two-index antisymmetric cubic operator in this model which will be marginal on the interface at one loop\cite{Osborn:2017ucf}. However, unlike the $O(4)$ model where such an operator leads to a manifold of fixed points at one loop, one finds by explicit computation that this operator does not satisfy $h_{ilm}h_{jln}h_{kmn}=0$ and thus will not generate fixed points of \eqref{eq:betaone}.

\subsection{Stability matrix}
The anomalous dimensions of cubic operators at a given fixed point are given by the eigenvalues, $\kappa$, of the stability matrix
\begin{equation}
    S_{ijk,lmn}=\frac{\partial\beta_{ijk}}{\partial h_{lmn}}
\end{equation}
evaluated at the fixed point. Relevant operators correspond to $\kappa<0$ and irrelevant operators correspond to $\kappa>0$. The breaking of a continuous bulk symmetry group $G$ to a subgroup $H<G$ by the introduction of the defect is associated with the existence of marginal $\kappa=0$ eigenvalues. These arise due to the eigenvector
\begin{equation}
    v_{ijk}=\omega_{ia}h_{ajk}+\omega_{ja}h_{aik}+\omega_{ka}h_{aij}
\end{equation}
of the stability matrix. Here $\omega_{ij}=-\omega_{ji}$ and
\begin{equation}
    \omega_{ia}\lambda_{ajkl}+\omega_{ja}\lambda_{iakl}+\omega_{ka}\lambda_{ijal}+\omega_{la}\lambda_{ijka}=0
\end{equation}
since $\lambda_{ijkl}$ is invariant under the action of $G$.

Examining the form of $v_{ijk}$, one sees that if $\omega_{ij}$ arises due to the remaining symmetry group $H$ then $v_{ijk}=0$, so that the statement $S_{ijk,lmn}\lsp v_{lmn}=0$ is only non-trivial when the $\omega_{ij}$ are chosen to correspond to the generators of $G$ broken by the defect. The number of $\kappa=0$ eigenvalues will then equal the number of generators of the bulk symmetry group broken at the defect fixed point, allowing one to easily determine the dimension of the remaining symmetry group $H$.

\section{Numerical results for low values of \texorpdfstring{$N$}{N}}
\label{sec:numerics}
For $N$ bulk scalar fields there will be a total of $N(N+1)(N+2)/3!$ independent defect beta functions one must solve in order to find fixed points. As $N$ increases, the analytic problem becomes intractable unless we assume that the defect perturbation takes a specific form to reduce the number of independent equations. When concerned with general questions about the symmetry breaking patterns which may arise at values of $N$ above two we must turn to numerical methods for answers. First, let us note that in extensive searches of coupling space it is essential to determine whether or not two fixed points are really distinct. The bulk tensor $\lambda_{ijkl}$ will be invariant under some bulk symmetry group $G\leq O(N)$, acting via rotation on the vector indices
\begin{equation}
    R_{ia}R_{jb}R_{kc}R_{ld}\lambda_{abcd}=\lambda_{ijkl}\,,\qquad R\in G\,.
\end{equation}
The introduction of a non-zero defect coupling $h_{ijk}$ will break $G$ to some defect symmetry group $H\leq G$. If we consider a rotation $R$ which is in $G$ but not in $H$, one can see that if $h_{1,ijk}$ is a fixed point, then so too will be
\begin{equation}
    h_{2,ijk}=R_{ia}R_{jb}R_{kc}h_{1,abc}\,.
\end{equation}
As the action of $G$ on the scalar fields $\phi_i$ is a field redefinition $\phi_i\rightarrow \phi'_i=R_{ij}\phi_j$, the fixed points $h_{1,ijk}$ and $h_{2,ijk}$ must be physically equivalent, so that the coupling space is properly defined modulo the action of $G$. It is thus more useful to work not directly with the fixed points $h_{ijk}$ one finds, but instead with the two $O(N)$ invariants
\begin{equation}
    h^2=h_{ijk}h_{ijk}\,,\qquad h_i^2=h_{ijj}h_{ikk}\,,
\end{equation}
which are the same for each member of the orbit of $h_{ijk}$ under $G$. It is with these invariants that we will characterise the fixed points we find.

To perform the numerical search itself we rely upon \textit{Mathematica}'s \texttt{FindRoot} function. As our initial input we take random sampling of points, with $0\leq h_{ijk}\leq1$ for all $i,j,k$. To verify the convergence, we take only those solutions with error less than $10^{-10}$ and then demand that at higher precision an improved solution can be found with error less than $10^{-100}$.

\subsection{\texorpdfstring{$N\leq3$}{N<=3}}
Before considering the case of interacting bulk models, let us consider an interface placed inside of a bulk containing $N$ free scalars. Unlike in the case of line defects, the existence of the purely defect term $T$ in the beta function permits the existence of non-trivial solutions to $\beta_{ijk}=0$ whilst setting $\lambda_{ijkl}=0$. We find no solutions for a free bulk with $N\leq 2$, while we find one non-trivial fixed point arising at $N=3$ given in Table \ref{tab:free3}.

\begin{center}
\begin{longtable}{|c c c c|}\caption{Fixed points found for an $N=3$ free bulk.} \\
\hline
Symmetry & $h^2$ & $h_i^2$ &  $\#\,\kappa<0,$ =0 \\ [0.5ex]
 \hline
\endfirsthead

\multicolumn{4}{c}%
{{\bfseries \tablename\ \thetable{} -- continued from previous page}} \\
\hline
Symmetry & $h^2$ & $h_i^2$ &  $\#\,\kappa<0,$ =0 \\ [0.5ex]
 \hline
\endhead

\hline \multicolumn{4}{|r|}{{Continued on next page}} \\ \hline
\endfoot

\hline
\endlastfoot \label{tab:free3}
$O(2)$ & 72.5147 & 6.1402 & 7, 2 \\
\end{longtable}
\end{center}
As indicated by the two $\kappa=0$ eigenvalues, the $O(3)$ bulk symmetry is broken to an $O(2)$ symmetry on the interface, so that the fixed point lies within the $O(N-1)$ symmetry breaking family studied previously for a free bulk in \cite{Harribey:2023xyv}.

Let us now take $\lambda_{ijkl}\neq0$. For low values of $N$ there are few fully interacting bulk critical models from which we can choose\cite{Osborn:2020cnf}, and it is thus possible to engage in a complete classification of critical interface models with only a few scalar fields. For $N=1$, the only interacting bulk point is the $\mathbb{Z}_2$ Wilson--Fisher fixed point, with
\begin{equation}
    \lambda_{1111}=\frac{\varepsilon}{3}\,.
\end{equation}
The beta function for the sole defect coupling is then given by
\begin{equation}
    \beta_{111}=\frac{\varepsilon}{2}h_{111}-\frac{1}{4}h_{111}^3\,,
\end{equation}
which has the unique non-trivial solution $h_{111}=\pm\sqrt{2\varepsilon}$ listed in Table \ref{tab:O1}. The introduction of the defect will completely break the $\mathbb{Z}_2$ symmetry in this system, reflected by the ambiguity in the choice of sign. Here, the stability matrix is simply given by the real number
\begin{equation}
    S=\frac{\varepsilon}{2}-\frac{3}{4}h_{111}^2=-\varepsilon\,.
\end{equation}
As this number is negative, the corresponding cubic operator will be relevant, and the fixed point will thus be unstable.
\begin{center}
\begin{longtable}{|c c c c|}\caption{Fixed points found for a $\mathbb{Z}_2$ bulk.} \\
\hline
Symmetry & $h^2$ & $h_i^2$ &  $\#\,\kappa<0,$ =0 \\ [0.5ex]
 \hline
\endfirsthead

\multicolumn{4}{c}%
{{\bfseries \tablename\ \thetable{} -- continued from previous page}} \\
\hline
Symmetry & $h^2$ & $h_i^2$ &  $\#\,\kappa<0,$ =0 \\ [0.5ex]
 \hline
\endhead

\hline \multicolumn{4}{|r|}{{Continued on next page}} \\ \hline
\endfoot

\hline
\endlastfoot \label{tab:O1}
None & 2 & 2 & 1, 0 \\
\end{longtable}
\end{center}

For $N=2$, there are now three choices of bulk critical models: a decoupled pair of Ising theories $I\times I$, a decoupled $I\times\text{Free}$ theory, and the $O(2)$ critical model. Let us first consider the former of these, where the interaction tensor can be expressed by
\begin{equation}
    \lambda_{ijkl}\phi_i\phi_j\phi_k\phi_l=\frac{\varepsilon}{3}\left(\phi_1^4+\phi_2^4\right),
\end{equation}
which is invariant under a symmetry group $\mathbb{Z}_2^2\rtimes\mathbb{Z}_2$. We now have four independent defect couplings,
\begin{equation}
    k_1=h_{111} \, ,\quad k_2=h_{222} \, , \quad g_1=h_{122} \, , \quad g_2=h_{112} \, ,
\end{equation}
the beta functions for which are
\begin{align}
    \beta_{k_1}&= \frac{\varepsilon}{2}k_1-\frac{k_1^3}{4}-\frac{g_1^3}{4}-\frac{3}{4}g_2^2\left(k_1+g_1\right) , \crcr
     \beta_{k_2}&= \frac{\varepsilon}{2}k_2-\frac{k_2^3}{4}-\frac{g_2^3}{4}-\frac{3}{4}g_1^2\left(k_2+g_2\right) , \crcr
     \beta_{g_1}&=-\frac{\varepsilon}{6}g_1-\frac{1}{4}\left(k_1+g_1\right)\left(g_1^2+g_2^2\right)-\frac{g_1}{4}\left(k_2+g_2\right)^2 , \crcr
      \beta_{g_2}&=-\frac{\varepsilon}{6}g_2-\frac{1}{4}\left(k_2+g_2\right)\left(g_1^2+g_2^2\right)-\frac{g_2}{4}\left(k_1+g_1\right)^2.
\end{align}
One can explicitly solve these equations to find that the only solutions are decoupled theories obtained via direct sums of the solution in Table \ref{tab:O1}, as shown in Table \ref{tab:IxI}.

\begin{center}
\begin{longtable}{|c c c c|}\caption{Fixed points found for a decoupled $I\times I$ bulk.} \\
\hline
Symmetry & $h^2$ & $h_i^2$ &  $\#\,\kappa<0,$ =0 \\ [0.5ex]
 \hline
\endfirsthead

\multicolumn{4}{c}%
{{\bfseries \tablename\ \thetable{} -- continued from previous page}} \\
\hline
Symmetry & $h^2$ & $h_i^2$ &  $\#\,\kappa<0,$ =0 \\ [0.5ex]
 \hline
\endhead

\hline \multicolumn{4}{|r|}{{Continued on next page}} \\ \hline
\endfoot

\hline
\endlastfoot \label{tab:IxI}
$\mathbb{Z}_2$ & 2 & 2 & 3, 0 \\
$\mathbb{Z}_2$ & 4 & 4 & 4, 0 \\
\end{longtable}
\end{center}

Taking the bulk to lie at the decoupled $I\times\text{Free}$ fixed point with symmetry $\mathbb{Z}_2^2$, where the interaction tensor is
\begin{equation}
    \lambda_{ijkl}\phi_i\phi_j\phi_k\phi_l=\frac{\varepsilon}{3}\lsp\phi_1^4\,,
\end{equation}
we find the four beta functions
\begin{align}
    \beta_{k_1}&= \frac{\varepsilon}{2}k_1-\frac{k_1^3}{4}-\frac{g_1^3}{4}-\frac{3}{4}g_2^2\left(k_1+g_1\right) , \crcr
     \beta_{k_2}&=-\frac{\varepsilon}{2}k_2-\frac{k_2^3}{4}-\frac{g_2^3}{4}-\frac{3}{4}g_1^2\left(k_2+g_2\right) , \crcr
     \beta_{g_1}&=-\frac{\varepsilon}{2}g_1-\frac{1}{4}\left(k_1+g_1\right)\left(g_1^2+g_2^2\right)-\frac{g_1}{4}\left(k_2+g_2\right)^2 , \crcr
      \beta_{g_2}&=-\frac{\varepsilon}{6}g_2-\frac{1}{4}\left(k_2+g_2\right)\left(g_1^2+g_2^2\right)-\frac{g_2}{4}\left(k_1+g_1\right)^2.
\end{align}
These may be explicitly solved, and as for the $I\times I$ theory the only resulting fixed points are obtained by direct sums of solutions for the Ising model and the $N=1$ free theory. As the $N=1$ free theory has no non-trivial defects, we are left with a single fixed point as shown in Table \ref{tab:IxF}.

\begin{center}
\begin{longtable}{|c c c c|}\caption{Fixed points found for an $I\times \text{Free}$ bulk.} \\
\hline
Symmetry & $h^2$ & $h_i^2$ &  $\#\,\kappa<0,$ =0 \\ [0.5ex]
 \hline
\endfirsthead

\multicolumn{4}{c}%
{{\bfseries \tablename\ \thetable{} -- continued from previous page}} \\
\hline
Symmetry & $h^2$ & $h_i^2$ &  $\#\,\kappa<0,$ =0 \\ [0.5ex]
 \hline
\endhead

\hline \multicolumn{4}{|r|}{{Continued on next page}} \\ \hline
\endfoot

\hline
\endlastfoot \label{tab:IxF}
$\mathbb{Z}_2$ & 2 & 2 & 4, 0 \\
\end{longtable}
\end{center}

Considering now the $O(2)$ model, the bulk interaction tensor takes the form
\begin{equation}
    \lambda_{ijkl}=\frac{\varepsilon}{10}\left(\delta_{ij}\delta_{kl}+\delta_{ik}\delta_{jl}+\delta_{il}\delta_{jk}\right).
    \label{eq:lambdaO2}
\end{equation}
The beta functions for the four independent couplings $k_1$, $k_2$, $g_1$ and $g_2$ are given by
\begin{align}
    \beta_{k_1}&= \frac{3\varepsilon}{10}k_1-\frac{k_1^3}{4}-\frac{g_1^3}{4}-\frac{3}{4}g_2^2\left(k_1+g_1\right) , \crcr
     \beta_{k_2}&= \frac{3\varepsilon}{10}k_2-\frac{k_2^3}{4}-\frac{g_2^3}{4}-\frac{3}{4}g_1^2\left(k_2+g_2\right) , \crcr
     \beta_{g_1}&=\frac{3\varepsilon}{10}g_1-\frac{1}{4}\left(k_1+g_1\right)\left(g_1^2+g_2^2\right)-\frac{g_1}{4}\left(k_2+g_2\right)^2 , \crcr
      \beta_{g_2}&=\frac{3\varepsilon}{10}g_2-\frac{1}{4}\left(k_2+g_2\right)\left(g_1^2+g_2^2\right)-\frac{g_2}{4}\left(k_1+g_1\right)^2.
\end{align}
These expressions may be solved numerically to yield the three non-trivial fixed points listed in Table \ref{tab:O2}. Importantly, it is possible to use the bulk $O(2)$ symmetry to rewrite the deformation $h_{ijk}$ at these fixed points in the simple form
\begin{equation}
    h_{ijk}\phi_i\phi_j\phi_k = 3\lsp\lambda_1\lsp\phi_2\phi_1^2+\lambda_2\lsp\phi_2^3\,,
\end{equation}
so that these fixed points fall into the family of fixed points classified previously in \cite{Harribey:2023xyv}. As one can see from the form of the potential and the number of $\kappa=0$ eigenvalues, these fixed points break the bulk $O(2)$ symmetry to a $\mathbb{Z}_2$ subgroup. At each of these fixed points the stability matrix has at least one negative eigenvalue, indicating that none of them are stable.

\begin{center}
\begin{longtable}{|c c c c|}\caption{Fixed points found for an $O(2)$ bulk.} \\
\hline
Symmetry & $h^2$ & $h_i^2$ & $\#\,\kappa<0$, =0 \\ [0.5ex]
 \hline
\endfirsthead

\multicolumn{4}{c}%
{{\bfseries \tablename\ \thetable{} -- continued from previous page}} \\
\hline
Symmetry & $h^2$ & $h_i^2$ &  $\#\,\kappa<0$, =0 \\ [0.5ex]
 \hline
\endhead

\hline \multicolumn{4}{|r|}{{Continued on next page}} \\ \hline
\endfoot

\hline
\endlastfoot \label{tab:O2}
$\mathbb{Z}_2$ & 1.7163 & 0.4849 & 1, 1 \\
$\mathbb{Z}_2$ & 2.7224 & 0.5823 & 2, 1\\
$\mathbb{Z}_2$ & 3.5613 & 4.5327  & 3, 1\\
\end{longtable}
\end{center}

For $N=3$ the situation becomes more complicated, with decoupled theories $I\times I\times I$, $O(2)\times I$, $I\times I\times\text{Free}$, $I\times \text{Free}\times\text{Free}$ and $O(2)\times\text{Free}$ and three interacting bulks one may consider: the $O(3)$ critical model, the cubic critical model with symmetry $B_3=\mathbb{Z}_2^3\rtimes S_3$, and the interacting $O(2)\times\mathbb{Z}_2$ biconical model. Let us first consider the situation with fully interacting bulks. Taking the bulk to lie at the $O(3)$ fixed point, the interaction tensor is given by
\begin{equation}
        \lambda_{ijkl}=\frac{\varepsilon}{11}\left(\delta_{ij}\delta_{kl}+\delta_{ik}\delta_{jl}+\delta_{il}\delta_{jk}\right).
\end{equation}
For $N=3$ there will be a total of ten beta functions, with numerical searches uncovering the seven fixed points listed in Table \ref{tab:O3}. Examining the number of $\kappa=0$ eigenvalues, one sees that three of these fixed points break the bulk $O(3)$ to an $O(2)$ symmetry on the defect, matching the three points found for $N=3$ in \cite{Harribey:2023xyv}. The remaining four points break the bulk $O(3)$ to a discrete subgroup. We can identify the remaining symmetry at one of these points by noting that they correspond to the $S_4$ fixed point constructed in section \ref{sec:ONbulk}. For the remaining three fixed points, we attempt to identify the symmetry by explicitly constructing the action of $O(3)$ on the tensor $h_{ijk}$ and identifying the stabiliser of a representative solution. Using the expression for a generic $O(3)$ rotation in terms of three angles,
\begin{equation}
    R_{ij}=\begin{pmatrix}
        \cos{\alpha}\cos{\beta} & -\cos{\gamma}\sin{\alpha}+\cos{\alpha}\sin{\beta}\sin{\gamma} & \cos{\alpha}\cos{\gamma}\sin{\beta}+\sin{\alpha}\sin{\gamma} \\
        \sin{\alpha}\cos{\beta} & \cos{\gamma}\cos{\alpha}+\sin{\alpha}\sin{\beta}\sin{\gamma} & \sin{\alpha}\cos{\gamma}\sin{\beta}-\cos{\alpha}\sin{\gamma} \\
        -\sin{\beta} & \cos{\beta}\sin{\gamma} & \cos{\beta}\cos{\gamma}
    \end{pmatrix},
\end{equation}
we numerically solve for $\alpha$, $\beta$ and $\gamma$ satisfying
\begin{equation}
    R_{ia}R_{jb}R_{kc}h_{abc}=h_{ijk}\,.
\end{equation}
For one fixed point, we find that there are six invariant rotations, forming a remaining symmetry group $S_3$. While it seems that a $\mathbb{Z}_2^2$ symmetry remains at the two other points, this is at odds with the existence of pairs of degenerate stability matrix eigenvalues at these points. These pairs of eigenvalues would normally suggest that the symmetry group ought to possess a two-dimensional irreducible representation, which is not the case for $\mathbb{Z}_2^2$. It is possible that the degeneracy of these eigenvalues is an accidental one loop effect which will be lifted upon the inclusion of higher order terms in the beta function.

\begin{center}
\begin{longtable}{|c c c c|}\caption{Fixed points found for an $O(3)$ bulk.} \\
\hline
Symmetry & $h^2$ & $h_i^2$ &  $\#\,\kappa<0$, =0 \\ [0.5ex]
 \hline
\endfirsthead

\multicolumn{4}{c}%
{{\bfseries \tablename\ \thetable{} -- continued from previous page}} \\
\hline
Symmetry & $h^2$ & $h_i^2$ & $\#\,\kappa<0$, =0 \\ [0.5ex]
 \hline
\endhead

\hline \multicolumn{4}{|r|}{{Continued on next page}} \\ \hline
\endfoot

\hline
\endlastfoot \label{tab:O3}
$O(2)$ & 0.9649 & 0.0491  & 3, 2 \\
$S_4$ & 1.909 & 0 & 1, 3 \\
$\mathbb{Z}_2^2$ & 3.574 & 0.8779 & 4, 3 \\
$\mathbb{Z}_2^2$ &  4.8467 & 2.1506 & 5, 3 \\
$O(2)$ & 5.1972 & 8.1514 & 8, 2\\
$O(2)$ & 6.0256 & 3.4477 & 7, 2\\
$S_3$ & 6.9037 & 3.1791 & 6, 3
\end{longtable}
\end{center}

The essential feature of Table \ref{tab:O3} is the variety of symmetry breaking patterns which arise, marking the difference between interface theories, and defects with lower dimension. For the $O(3)$ line defect theory, the only possible defect fixed point retained a continuous $O(2)$ symmetry corresponding to choosing a preferred direction\cite{Pannell:2023pwz}. While the $O(3)$ surface defect possesses one more fixed point than the $O(3)$ line defect, the remaining symmetry groups, $O(3)$ and $O(2)\times\mathbb{Z}_2$ respectively, are both continuous\cite{Trepanier:2023tvb}. With the interface, it is possible to break the symmetry generators entirely and gain access to the discrete subgroups of $O(3)$.

Switching now to the critical cubic bulk, the tensor $\lambda_{ijkl}$ has the form
\begin{equation}
        \lambda_{ijkl}=\frac{\varepsilon}{9}\left(\delta_{ij}\delta_{kl}+\delta_{ik}\delta_{jl}+\delta_{il}\delta_{jk}\right)-\frac{\varepsilon}{9}\lsp\delta_{ijkl}\,,
\end{equation}
where $\delta_{ijkl}$ is the four-index generalisation of the Kronecker delta. We again solve the resulting ten beta functions numerically to find a total of 26 distinct fixed points, shown in Table \ref{tab:B3}. As the initial bulk symmetry is discrete, breaking is not accompanied by any marginal tilt operators, which is reflected by the lack of $\kappa=0$ eigenvalues. Of note is the fact that each of the fixed points has at least one $\kappa<0$ eigenvalue, so that they are all unstable. Two of these points have negative-definite stability matrices, indicating that these will be UV-stable rather than IR-stable fixed points. The stability that remains at a fixed point can be determined in a number of different ways. As the cubic group is of order $8\times 3!=48$, the orbit of a fixed point, the set of fixed points with the same values of $h^2$ and $h_i^2$, must be an integer divisor of 48. By the orbit stabiliser theorem, the order of the remaining symmetry group $H$ can thus be determined provided the numerical computations are able to generate the complete orbit. In practice the numerical solver does not seem to be able to generate such complete lists and often the order of the orbit one finds is not a divisor of 48. Instead, we will determine the symmetry by acting explicitly on a representative member of an orbit with the elements of $B_3$. As the cubic group is finite, this is easier than in the $O(3)$ case, and we are able to determine precisely which elements preserve the fixed point tensor $h_{ijk}$. The result is displayed in the first column of Table \ref{tab:B3}.

\begin{center}
\begin{longtable}{|c c c c|}\caption{Fixed points found for a cubic bulk.} \\
\hline
Symmetry & $h^2$ & $|h_i|^2$ & $\#\,\kappa<0$, =0 \\ [0.5ex]
 \hline
\endfirsthead

\multicolumn{4}{c}%
{{\bfseries \tablename\ \thetable{} -- continued from previous page}} \\
\hline
Symmetry & $h^2$ & $|h_i|^2$ & $\#\,\kappa<0$, =0 \\ [0.5ex]
 \hline
\endhead

\hline \multicolumn{4}{|r|}{{Continued on next page}} \\ \hline
\endfoot

\hline
\endlastfoot \label{tab:B3}
$\mathbb{Z}_2$ & 1.5652 & 0.1391 & 5, 0 \\
$\mathbb{Z}_2^2$ & 3.153 & 0.3772 & 5, 0 \\
$\mathbb{Z}_2^2$ & 3.242 & 1.1784 & 6, 0 \\
$S_3$ & 3.5151 & 0.2 & 6, 0 \\
$\mathbb{Z}_2$ & 3.8128 & 1.0632 & 5, 0 \\
$S_3$ & 3.9508 & 2.7339 & 6, 0 \\
$S_4$ & 4 & 0 & 7, 0 \\
$\mathbb{Z}_2$ & 4.0567 & 0.9064 & 6, 0 \\
$\mathbb{Z}_2$ & 4.0649 & 1.0211 & 7, 0 \\
None & 4.1977 & 1.8209 & 6, 0 \\
None & 4.3628 & 1.7113 & 7, 0 \\
$\mathbb{Z}_2$ & 4.3682 & 1.073 & 6, 0 \\
$\mathbb{Z}_2$ & 4.4154 & 1.6658 & 8, 0 \\
$\mathbb{Z}_2$ & 4.6101 & 2.5405 & 7, 0 \\
$S_3$ & 4.7554 & 2.4035 & 7, 0 \\
$\mathbb{Z}_2^2$ & 4.8298 & 7.5922 & 8, 0 \\
$\mathbb{Z}_2$ & 5.0071 & 7.1291 & 9, 0 \\
$S_3$ & 5.022 & 7.0846 & 10, 0 \\
$\mathbb{Z}_2$ & 5.1195 & 1.6451 & 7, 0 \\
$\mathbb{Z}_2^2$ & 5.3959 & 2.9045 & 8, 0 \\
$\mathbb{Z}_2$ & 5.6296 & 2.6667 & 7, 0 \\
$D_4$ & 5.8115 & 8.546 & 10, 0 \\
$\mathbb{Z}_2$ & 6.0861 & 2.7389 & 8, 0 \\
$\mathbb{Z}_2$ & 6.657 & 3.1942 & 8, 0 \\
$S_3$ & 7.1355 & 4.7822 & 9, 0 \\
$D_4$ & 7.6863 & 4.1457 & 9, 0 \\
\end{longtable}
\end{center}

One again sees that the symmetry breaking patterns which can appear are more intricate than for other types of defects. While the cubic surface defect has not yet been studied, the cubic line defect was investigated in \cite{Pannell:2023pwz}. Three non-trivial dCFTs were identified, with $D_4$, $\mathbb{Z}_2^2$ and $S_4$ symmetry corresponding to choosing the privileged vector to lie in the centre of a face, edge or vertex respectively. In Table \ref{tab:B3} these symmetry groups are not only joined by the groups $\mathbb{Z}_2$ and $S_3$, but one also sees that there may exist multiple realisations of the same symmetry group. Rather interestingly, it is also possible for the introduction of the interface to completely break the bulk symmetry, something which was not observed with the line defect.

Let us now consider taking the bulk to lie at the interacting $O(2)\times\mathbb{Z}_2$ biconical critical bulk point. Here, the interaction tensor has irrational coefficients, and it is best expressed by contracting with factors of $\phi_i$
\begin{equation}
    \lambda_{ijkl}\phi_i\phi_j\phi_k\phi_l=0.252845\lsp(\phi_1^2+\phi_2^2)^2+0.690589\lsp(\phi_1^2+\phi_2^2)\phi_3^2+0.20248\lsp\phi_3^4\,.
    \label{eq:lambdabiconical}
\end{equation}
Solving the beta functions numerically we find a total of nine fixed points, displayed in Table \ref{tab:O2I}. As may be expected from the irrational bulk couplings, all of the fixed points have irrational invariants. The single $\kappa=0$ eigenvalue at each fixed point indicates that in all cases the bulk $O(2)$ symmetry is broken by the defect to a discrete subgroup. As before, we determine the remaining symmetry by explicitly constructing the action of $O(2)\times\mathbb{Z}_2$ on a representative solution from each equivalency class. The result is displayed in the first column of Table \ref{tab:O2I}. It is also worth noting that as for the other two bulks, each fixed point possesses at least one relevant operator, and is thus unstable.

\begin{center}
\begin{longtable}{|c c c c|}\caption{Fixed points found for an $O(2)\times \mathbb{Z}_2$ biconical bulk.} \\
\hline
Symmetry & $h^2$ & $|h_i|^2$ & $\#\,\kappa<0$, =0 \\ [0.5ex]
 \hline
\endfirsthead

\multicolumn{4}{c}%
{{\bfseries \tablename\ \thetable{} -- continued from previous page}} \\
\hline
Symmetry & $h^2$ & $|h_i|^2$ & $\#\,\kappa<0$, =0 \\ [0.5ex]
 \hline
\endhead

\hline \multicolumn{4}{|r|}{{Continued on next page}} \\ \hline
\endfoot

\hline
\endlastfoot \label{tab:O2I}
$\mathbb{Z}_2^2$ & 0.12418 & 0.00220839 & 2, 1\\
$\mathbb{Z}_2^2$ & 0.143199 & 0.00306045 & 3, 1\\
$\mathbb{Z}_2^2$ & 1.43105 & 0.0813986 & 4, 1\\
$\mathbb{Z}_2$ & 2.45154 & 0.245134 & 4, 1\\
$D_4$ & 3.09495 & 0 & 6, 1 \\
$\mathbb{Z}_2$ & 3.53387 & 1.30149 & 5, 1 \\
$\mathbb{Z}_2^2$ & 5.40366 & 8.28185 & 9, 1 \\
$\mathbb{Z}_2^2$ & 6.66652 & 3.59745 & 8, 1 \\
$\mathbb{Z}_2^2$ & 7.93982 & 0.279441 & 5, 1
\end{longtable}
\end{center}

\begin{figure}[H]
\centering
\begin{tikzpicture}
\begin{axis}[
    xmin=-0, xmax=9,
    ymin=0, ymax=10,
    xlabel= $h^2$,
    ylabel= $|h_i|^2$,
    ylabel style={rotate=-90},
    title={Fixed points for $N=3$ fully interacting bulks},
    legend pos = outer north east,
    reverse legend,
]
\addplot+[
    only marks,
    mark=square*,
    mark options={color=Dark2-B,fill=Dark2-B},
    mark size=2pt]
table{Data/B3.dat};
\addlegendentry{Cubic bulk}
\addplot+[
    only marks,
    mark=triangle*,
    mark options={color=Dark2-C,fill=Dark2-C},
    mark size=2pt]
table{Data/BO2I.dat};
\addlegendentry{$O(2)\times \mathbb{Z}_2$ bulk}
\addplot+[
    only marks,
    mark=*,
    mark options={color=Dark2-A,fill=Dark2-A},
    mark size=2pt]
table{Data/O3.dat};
\addlegendentry{$O(3)$ bulk}
\end{axis}
\end{tikzpicture}
    \caption{Unitary fixed points for $N=3$ interacting bulks found using a numerical search beginning with 10,000 points. We find 7 fixed points in an $O(3)$ bulk, 26 in a cubic bulk, and 9 in an $O(2)\times\mathbb{Z}_2$ biconical bulk. Not shown is the trivial fixed point $h_{ijk}=0$, which is the only stable point for all three bulks.}
\label{fig:Neq3interacting}
\end{figure}
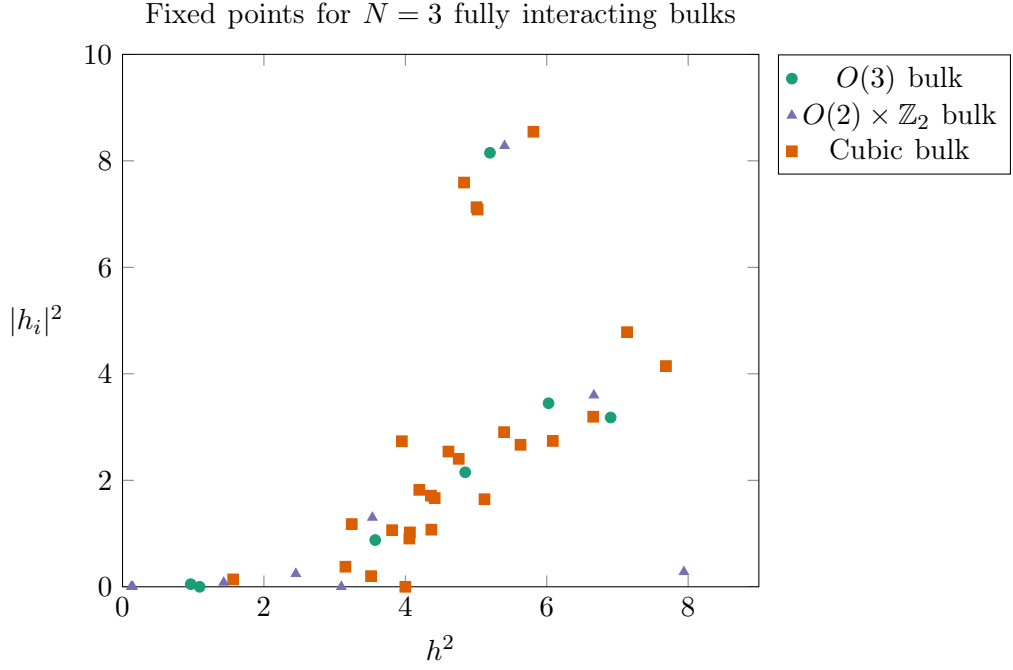

We display the results of Tables \ref{tab:O3}\,-\,\ref{tab:O2I} graphically in Fig.\;\ref{fig:Neq3interacting}, plotting the values of the $O(3)$ invariants $h_i^2$ versus $h^2$. As these three bulks are the only possible fully interacting critical points for three scalar fields, this set of fixed points will be a complete classification of interface dCFTs with three scalars in a critical interacting bulk theory. It is interesting to note that the fixed points of the various bulk models occupy similar regions of this space of invariants, perhaps suggesting that the form of the fixed point solutions are related.

We can now turn to decoupled bulk models. While one could restrict one's attention to deforming only by operators which do not combine fields not already coupled in the bulk, it is more interesting to consider fixed points which cause the bulk fields to become interacting on the defect. Let us first consider the bulk to lie at an $O(2)$ point for the fields $\phi_1$ and $\phi_2$, while the field $\phi_3$ is left free, giving the bulk a symmetry group of $O(2)\times\mathbb{Z}_2$. To be precise, the interaction tensor will take the form
\begin{equation}
    \lambda_{ijkl}\phi_i\phi_j\phi_k\phi_l=\frac{3\varepsilon}{10}\left(\phi_1^2+\phi_2^2\right)^2\,,
\end{equation}
where unlike (\ref{eq:lambdaO2}) $i,j,k,l$ run from 1 to 3. The numerical results are displayed in Table \ref{tab:OxF}. As there exist no non-trivial fixed points for the $N=1$ free bulk, we find only the three decoupled fixed points deriving from Table \ref{tab:O2}. The remaining three fixed points cause $\phi_3$ to non-trivially interact with $\phi_1$ and $\phi_2$ along the interface. One of these points retains the $O(2)$ rotational symmetry of $\phi_1$ and $\phi_2$, while the other two completely break the bulk symmetry group.

\begin{center}
\begin{longtable}{|c c c c|}\caption{Fixed points found for an $O(2)\times \text{Free}$ bulk.} \\
\hline
Symmetry & $h^2$ & $|h_i|^2$ & $\#\,\kappa<0$, =0 \\ [0.5ex]
 \hline
\endfirsthead

\multicolumn{4}{c}%
{{\bfseries \tablename\ \thetable{} -- continued from previous page}} \\
\hline
Symmetry & $h^2$ & $|h_i|^2$ & $\#\,\kappa<0$, =0 \\ [0.5ex]
 \hline
\endhead

\hline \multicolumn{4}{|r|}{{Continued on next page}} \\ \hline
\endfoot

\hline
\endlastfoot \label{tab:OxF}
$\mathbb{Z}_2^2$&1.7163& 0.4849&6, 1\\ $\mathbb{Z}_2^2$& 2.7224 & 0.5823&7, 1\\ $\mathbb{Z}_2^2$&3.5613& 4.5327&8, 1\\ $O(2)$ & 25.7941 & 2.8091 & 7, 0 \\ None & 27.2925 & 3.4681 & 7, 1 \\ None & 27.6885 & 3.6268 & 8, 1\\
\end{longtable}
\end{center}

We can also take the bulk to be the direct sum of an $O(2)$ theory and an Ising theory. Here, the interaction tensor may be expressed as
\begin{equation}
    \lambda_{ijkl}\phi_i\phi_j\phi_k\phi_l=\frac{3\varepsilon}{10}\left(\phi_1^2+\phi_2^2\right)^2+\frac{\varepsilon}{3}\phi_3^4\,,
\end{equation}
which differs from (\ref{eq:lambdabiconical}) in the lack of a term mixing $\phi_3$ with $\phi_1$ and $\phi_2$. Solving the beta functions numerically, we find that, unlike for the decoupled $I\times I$ theory, direct sums of $O(2)$ solutions with $\mathbb{Z}_2$ solutions are not the only possible fixed points. In the end a total of 12 fixed points appear, as displayed in Table \ref{tab:O2I}. Seven of these correspond to decoupled fixed points, whose symmetry may be easily determined from the form of the potential. The remaining five are theories in which the $\phi_3$ field, decoupled in the bulk, is able to interact along the interface with the other two fields $\phi_1$ and $\phi_2$. Examining the $\kappa=0$ eigenvalues, one sees that one of these extra points preserves the $O(2)$ symmetry of the $\phi_1$ and $\phi_2$ fields. To determine the remaining symmetry at the other fixed points we once again construct the action of $O(2)$ on the fields, and determine that no symmetry remains.
\begin{center}
\begin{longtable}{|c c c c|}\caption{Fixed points found for a decoupled $O(2)\times I$ bulk.} \\
\hline
Symmetry & $h^2$ & $|h_i|^2$ & $\#\,\kappa<0$, =0 \\ [0.5ex]
 \hline
\endfirsthead

\multicolumn{4}{c}%
{{\bfseries \tablename\ \thetable{} -- continued from previous page}} \\
\hline
Symmetry & $h^2$ & $|h_i|^2$ & $\#\,\kappa<0$, =0 \\ [0.5ex]
 \hline
\endhead

\hline \multicolumn{4}{|r|}{{Continued on next page}} \\ \hline
\endfoot

\hline
\endlastfoot \label{tab:O2Idecoupled}
$\mathbb{Z}_2^2$&1.7163& 0.4849&6, 1\\ $O(2)$&2& 2&6, 0\\ $\mathbb{Z}_2^2$& 2.7224 & 0.5823&7, 1\\ $O(2)$& 2.9215 & 0.4738&7, 0\\ $\mathbb{Z}_2^2$&3.5613& 4.5327&8, 1\\ $\mathbb{Z}_2$ & 3.7163 & 2.4849& 7, 1\\ $\mathbb{Z}_2$& 4.7224& 2.5823&8, 1 \\ $\mathbb{Z}_2$&5.5613&6.5327&9, 1\\ None &7.2701& 1.8644&7, 1\\ None&7.5703& 1.9352& 7, 1\\None& 9.4071& 1.7039&8, 1 \\ None &9.9475& 2.6177&8, 1
\end{longtable}
\end{center}

Let us now consider a bulk which is taken to be the direct sum of three decoupled Ising fixed points. The interaction tensor in this case will be
\begin{equation}
    \lambda_{ijkl}\phi_i\phi_j\phi_k\phi_l=\frac{\varepsilon}{3}\left(\phi_1^4+\phi_2^4+\phi_3^4\right),
\end{equation}
which is invariant under the cubic group $B_3=\mathbb{Z}_2^3\rtimes S_3$. There exists a fictitious conformal manifold in this model, given explicitly by the deformation
\begin{equation}
    3h\phi_1^2\phi_3+\frac{2}{h}\phi_2^2\phi_3-\frac{2+3h^2}{3h}\phi_3^3\,.
\end{equation}
One can check that these fixed points do not survive at two loops, and should thus be ignored in a complete classification. Noting that these fixed points have the same $O(N)$ invariant $h_i^2=0$ for all values of $h$, in the numerical search we allow only fixed points with $h_i^2\neq0$. While it is possible that there exist other, genuine fixed points with vanishing $h_i^2$, we note that in other decoupled bulks all fixed points obey $h_i^2\neq0$, so that we believe that our numerical search will still be exhaustive. In the end, the numerical search reveals the eight fixed points shown in Table \ref{tab:IxIxI}. As the bulk symmetry group is finite, it is possible to explicitly construct the broken generators to determine the remaining symmetry at these fixed points, as in the case of the cubic fixed point.

\begin{center}
\begin{longtable}{|c c c c|}\caption{Fixed points found for a decoupled $I\times I\times I$ bulk.} \\
\hline
Symmetry & $h^2$ & $|h_i|^2$ & $\#\,\kappa<0$, =0 \\ [0.5ex]
 \hline
\endfirsthead

\multicolumn{4}{c}%
{{\bfseries \tablename\ \thetable{} -- continued from previous page}} \\
\hline
Symmetry & $h^2$ & $|h_i|^2$ & $\#\,\kappa<0$, =0 \\ [0.5ex]
 \hline
\endhead

\hline \multicolumn{4}{|r|}{{Continued on next page}} \\ \hline
\endfoot

\hline
\endlastfoot \label{tab:IxIxI}
$\mathbb{Z}_2^3$&2& 2&8, 0\\
$\mathbb{Z}_2^2$ & 4 & 4 & 9, 0\\
$S_3$ & 6 & 6 & 10, 0\\
$\mathbb{Z}_2$ & 11.8304 & 0.9834 & 8, 0\\
$\mathbb{Z}_2$ & 13.4965 & 1.6436 & 7, 0\\
$S_3$ & 17.1538 & 3.0936 & 9, 0\\
$\mathbb{Z}_2$ & 20.2542 & 2.0717 & 8, 0\\
$\mathbb{Z}_2^2$ & 25.3646 & 0.4005 & 8, 0
\end{longtable}
\end{center}

Removing the $\phi_3^4$ term, we arrive at a bulk with two fields at the Ising point and one free field,
\begin{equation}
    \lambda_{ijkl}\phi_i\phi_j\phi_k\phi_l=\frac{\varepsilon}{3}\left(\phi_1^4+\phi_2^4\right),
\end{equation}
with a bulk symmetry group $(\mathbb{Z}_2^2\rtimes\mathbb{Z}_2)\times\mathbb{Z}_2$. The numerics produce the five fixed points in Table \ref{tab:IxIxF}, where once again because the symmetry group is finite we are able to explicitly determine the number of broken generators at each fixed point. Two of these points correspond to decoupled direct sums of the Ising interface with a free $\phi_3$, but the remaining points non-trivially couple the three fields.

\begin{center}
\begin{longtable}{|c c c c|}\caption{Fixed points found for an $I\times I\times \text{Free}$ bulk.} \\
\hline
Symmetry & $h^2$ & $|h_i|^2$ & $\#\,\kappa<0$, =0 \\ [0.5ex]
 \hline
\endfirsthead

\multicolumn{4}{c}%
{{\bfseries \tablename\ \thetable{} -- continued from previous page}} \\
\hline
Symmetry & $h^2$ & $|h_i|^2$ & $\#\,\kappa<0$, =0 \\ [0.5ex]
 \hline
\endhead

\hline \multicolumn{4}{|r|}{{Continued on next page}} \\ \hline
\endfoot

\hline
\endlastfoot \label{tab:IxIxF}
$\mathbb{Z}_2^2$&2& 2&9, 0\\
$\mathbb{Z}_2^2$ & 4 & 4 & 10, 0\\
$\mathbb{Z}_2$ & 33.5888 & 3.4233 & 7, 0\\
$\mathbb{Z}_2^2\rtimes\mathbb{Z}_2$ & 33.5912 & 3.3578 & 7, 0 \\
$\mathbb{Z}_2$ & 34.7502 & 3.7924 & 8, 0
\end{longtable}
\end{center}

Finally, we can take two of the bulk fields to be free, and the remaining field, which we choose to be $\phi_1$ to lie at the Ising critical point,
\begin{equation}
    \lambda_{ijkl}\phi_i\phi_j\phi_k\phi_l=\frac{\varepsilon}{3}\phi_1^4.
\end{equation}
This will have a bulk symmetry group $G=O(2)\times\mathbb{Z}_2$. We again find five fixed points, exhibited in Table \ref{tab:IxFxF}. The first of these points is decoupled, but the remaining four are fully interacting fixed points.

\begin{center}
\begin{longtable}{|c c c c|}\caption{Fixed points found for an $I\times \text{Free}\times \text{Free}$ bulk.} \\
\hline
Symmetry & $h^2$ & $|h_i|^2$ & $\#\,\kappa<0$, =0 \\ [0.5ex]
 \hline
\endfirsthead

\multicolumn{4}{c}%
{{\bfseries \tablename\ \thetable{} -- continued from previous page}} \\
\hline
Symmetry & $h^2$ & $|h_i|^2$ & $\#\,\kappa<0$, =0 \\ [0.5ex]
 \hline
\endhead

\hline \multicolumn{4}{|r|}{{Continued on next page}} \\ \hline
\endfoot

\hline
\endlastfoot \label{tab:IxFxF}
$O(2)$&2& 2&10, 0\\
$O(2)$ & 44.3429 & 2.3688 & 9, 0\\
None & 53.0497 & 4.5056 & 7, 1\\
$\mathbb{Z}_2$ & 53.0504 & 4.4683 & 7, 1\\
None & 54.1261 & 4.7865 & 8, 1
\end{longtable}
\end{center}

\begin{figure}[H]
\centering
\begin{tikzpicture}
\begin{axis}[
    xmin=-0, xmax=80,
    ymin=0, ymax=10,
    xlabel= $h^2$,
    ylabel= $|h_i|^2$,
    ylabel style={rotate=-90},
    title={Fixed points for $N=3$ decoupled bulks},
    legend pos = outer north east,
]
\addplot+[
    only marks,
    mark=Mercedes star,
    mark options={color=black},
    mark size=3pt]
table{Data/Free3.dat};
\addlegendentry{Free bulk}
\addplot+[
    only marks,
    mark=*,
    mark options={color=Dark2-A},
    mark size=2pt]
table{Data/O2xF.dat};
\addlegendentry{$O(2)\times \text{Free}$ Bulk}
\addplot+[
    only marks,
    mark=triangle*,
    mark options={color=Dark2-C},
    mark size=2pt]
table{Data/O2xI.dat};
\addlegendentry{$O(2)\times I$ Bulk}
\addplot+[
    only marks,
    mark=square*,
    mark options={color=Dark2-B},
    mark size=2pt]
table{Data/IxIxI.dat};
\addlegendentry{$I\times I\times I$ Bulk}
\addplot+[
    only marks,
    mark=diamond*,
    mark options={color=black},
    mark size=2pt]
table{Data/IxIxF.dat};
\addlegendentry{$I\times I\times\text{Free}$ Bulk}
\addplot+[
    only marks,
    mark=10-pointed star,
    mark options={color=black},
    mark size=3pt]
table{Data/IxFxF.dat};
\addlegendentry{$I\times\text{Free}\times\text{Free}$ Bulk}
\end{axis}
\end{tikzpicture}
    \caption{Unitary fixed points for $N=3$ decoupled bulks found using a numerical search beginning with 10,000 points. The fixed points can also be found in Table \ref{tab:free3} and Tables \ref{tab:OxF}\,-\ref{tab:IxFxF}.}
    \label{fig:Neq3decoupled}
\end{figure}
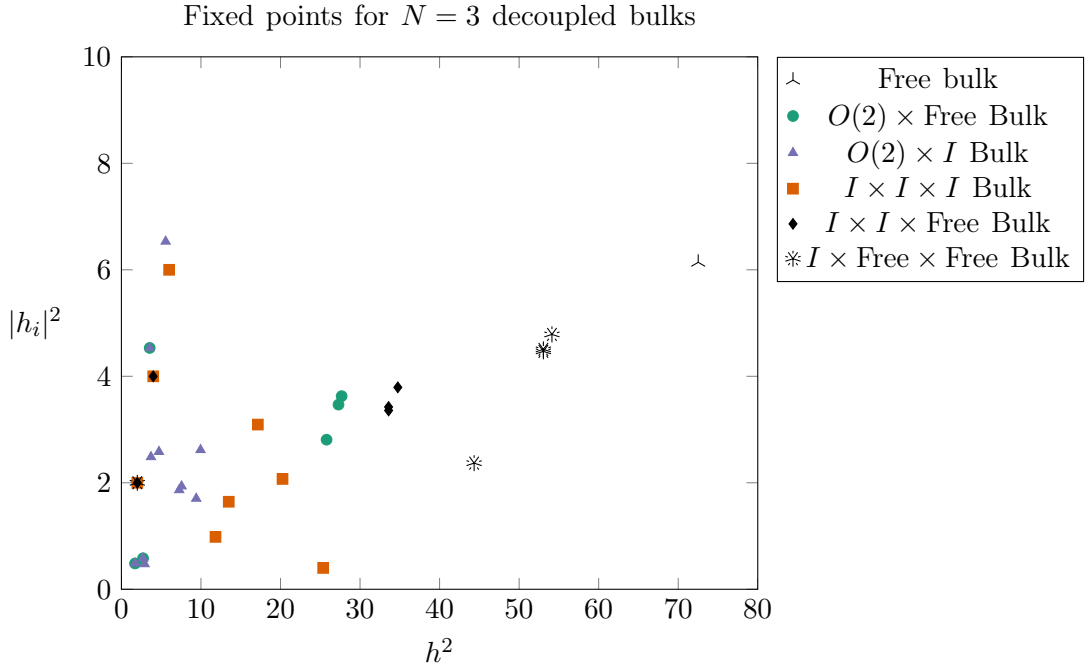

We display the results of Table \ref{tab:free3} and Tables \ref{tab:O2Idecoupled}\,-\ref{tab:IxFxF} in Fig.\;\ref{fig:Neq3decoupled}. Comparing with Fig.\;\ref{fig:Neq3interacting} one notices immediately that while the decoupled bulks can have considerably larger values of $h^2$, all fixed points seem to obey a bound of the form $|h_i|^2\lesssim 9$. As we have considered all possible bulks for $N\leq3$, the fixed points we have identified should provide a complete classification of scalar interface dCFTs with three or fewer fields.

\subsection{Larger numbers of fields}
For larger numbers of fields the number of possible critical bulk models one can choose from grows rapidly, and we will thus focus only on placing interfaces within well-known models. As our bulk models we choose the free bulk, the $O(N)$ model, the hypercubic $B_N=\mathbb{Z}_2^N\rtimes S_N$ model and the two hypertetrahedral $T_N^{\pm}=S_{N+1}\times \mathbb{Z}_2$ models, which have applications in a variety of physical systems. The number of fixed points we find numerically grows rapidly with $N$, as indicated by Table \ref{tab:numbers}. Here we have used a number of small $N$ isomorphisms between bulk fixed points and results from Tables \ref{tab:free3}-\ref{tab:IxFxF} to include results for hypercubic and hypertetrahedral models even for $N$
\begin{center}
\begin{longtable}{|c | c c c c c c c c|}\caption{Number of fixed points found for various $N$} \\
\hline
Bulk Model & $N=1$ & $N=2$ & $N=3$ & $N=4$ & $N=5$ & $N=6$ & $N=7$ & $N=8$\\ [0.5ex]
 \hline
\endfirsthead

\multicolumn{9}{c}%
{{\bfseries \tablename\ \thetable{} -- continued from previous page}} \\
\hline
Bulk Model & $N=1$ & $N=2$ & $N=3$ & $N=4$ & $N=5$ & $N=6$ & $N=7$ & $N=8$\\ [0.5ex]
 \hline
\endhead

\hline \multicolumn{4}{|r|}{{Continued on next page}} \\ \hline
\endfoot

\hline
\endlastfoot \label{tab:numbers}
Free & 0 & 0 & 1 & 5 & 18 & 48 & 136 & 71 \\
$O(N)$ & 1 & 3 & 7 & N/A & 14 & 35 & 159 & 50\\
$B_N$& 1 & 2 & 26 & N/A & 883 & 616 & 293 & 16\\
$T_N^-$& 1 & 3 & 26 & 169 & 1671 & 1560 & 395 & 28\\
$T_N^+$& 1 & 0 & 8 & N/A & 1671 & 1584 & 350 & 23
\end{longtable}
\end{center}
in which they were not explicitly considered. N/A indicates the presence of a leading-order conformal manifold, like the one discussed at the end of section \ref{sec:ONbulk}, which prevented the numerics from providing accurate results. Though the number of fixed points found decreases sharply for the larger $N$ we looked at, we suspect that this is a numerical artefact associated with \textit{Mathematica} facing difficulty working within a high-dimensional solution space. All of the numerical results were derived beginning with 10,000 initial points, and it is likely that at higher $N$ this is not sufficient to provide a satisfactory search of coupling space. Our numerical results stop at $N=8$, where there are already 120 equations and unknowns, as beyond this point the \texttt{FindRoot} function becomes prohibitively slow.

\begin{figure}[H]
    \centering
\begin{tikzpicture}
\begin{groupplot}[
    group style={
        group size=2 by 1,
        yticklabels at=edge left,
        horizontal sep=0pt,
        xlabels at=edge bottom,
        ylabels at=edge left
    },
    ymin=0, ymax=16,
    ylabel= $|h_i|^2$,
    ylabel style={rotate=-90},height=10cm,width=10cm
]

\nextgroupplot[xmin=0,xmax=17,
               xtick={0,2,4,6,8,10,12,14,16},
               width=6.75cm,
               height=9cm,
               axis y line=left,
               xlabel= $h^2$,
               xlabel style={at={(0.8, -0.06)}},
               title={Fixed points for $N=4$},
               title style={at={(0.8, 1.02)}}, ]
\addplot+[
    only marks,
    mark=diamond*,
    mark options={color=Dark2-F,fill=Dark2-F},
    mark size=1.5pt]
table{Data/T4m.dat};

\nextgroupplot[xmin=55,xmax=80,
               axis x discontinuity=parallel,
               xtick={60,65,70,75},
               width=4.5cm,
               height=9cm,
               axis y line = right,
               legend pos = outer north east,
               reverse legend,]
\addplot+[
    only marks,
    mark=diamond*,
    mark options={color=Dark2-F,fill=Dark2-F},
    mark size=1.5pt]
table{Data/T4m.dat};
\addlegendentry{$T_4^-$ bulk}
\addplot+[
    only marks,
    mark=Mercedes star,
    mark options={color=black},
    mark size=3pt]
table{Data/Free4.dat};
\addlegendentry{Free bulk}
\end{groupplot}
\end{tikzpicture}
    \caption{Unitary fixed points for $N=4$ found using a numerical search beginning with 10,000 points.}
 \label{fig:Neq4}
\end{figure}
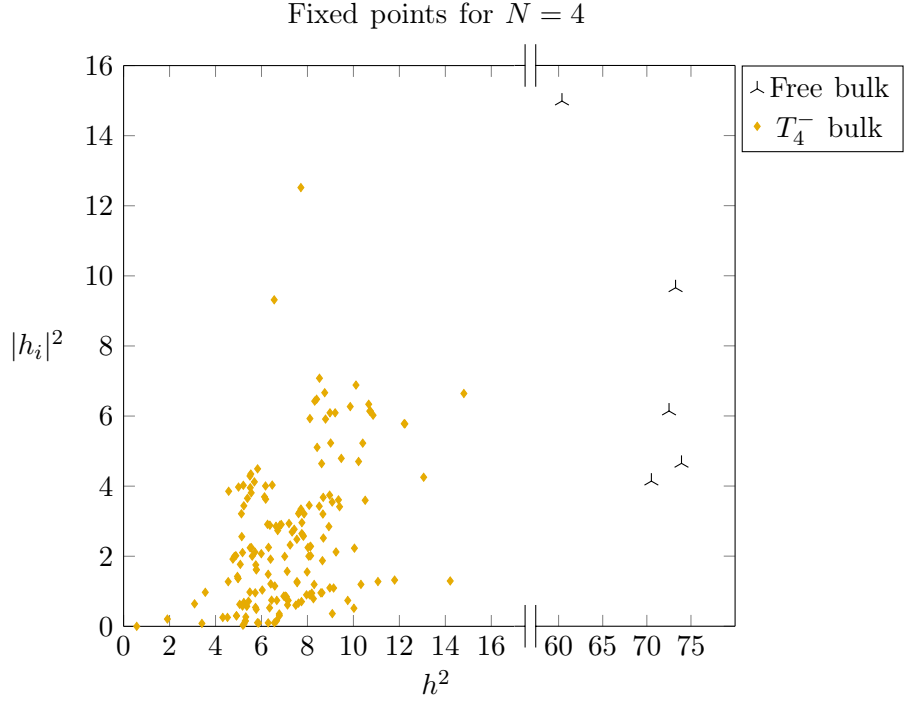

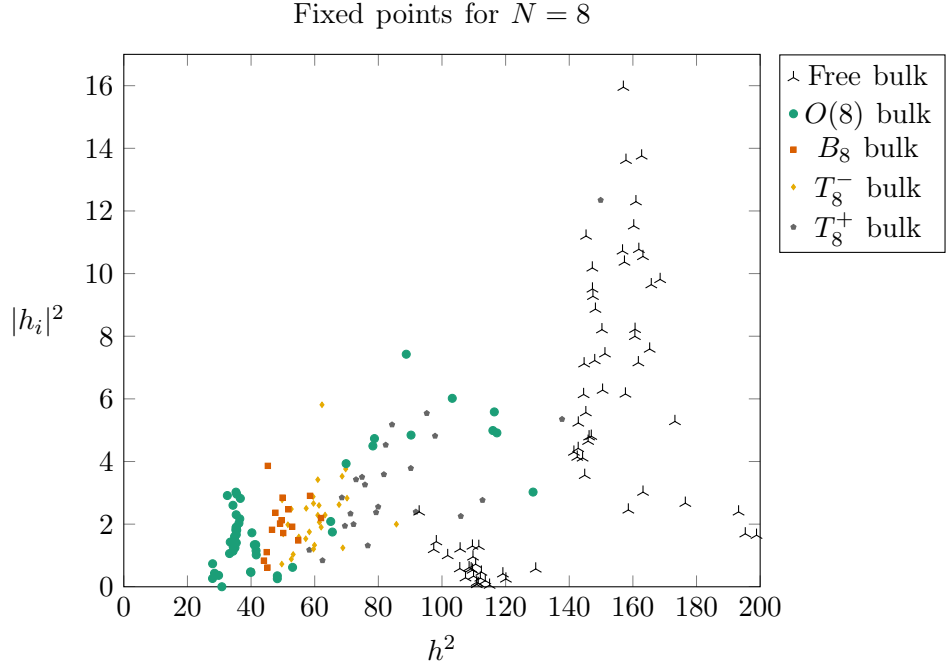
\begin{figure}[H]
\centering
\begin{tikzpicture}
\begin{axis}[
    xmin=-0, xmax=200,
    ymin=0, ymax=17,
    xlabel= $h^2$,
    ylabel= $|h_i|^2$,
    ylabel style={rotate=-90},
    title={Fixed points for $N=8$},
    legend pos = outer north east,
    reverse legend,
]
\addplot+[
    only marks,
    mark=pentagon*,
    mark options={color=Dark2-H,fill=Dark2-H},
    mark size=1pt]
table{Data/T8p.dat};
\addplot+[
    only marks,
    mark=diamond*,
    mark options={color=Dark2-F,fill=Dark2-F},
    mark size=1pt]
table{Data/T8m.dat};
\addplot+[
    only marks,
    mark=square*,
    mark options={color=Dark2-B,fill=Dark2-B},
    mark size=1pt]
table{Data/B8.dat};
\addplot+[
    only marks,
    mark=*,
    mark options={color=Dark2-A,fill=Dark2-A},
    mark size=1.5pt]
table{Data/O8.dat};
\addplot+[
    only marks,
    mark=Mercedes star,
    mark options={color=black},
    mark size=2.5pt]
table{Data/Free8.dat};
\legend{$T_8^+$ bulk,$T_8^-$ bulk,$B_8$ bulk,$O(8)$ bulk, Free bulk}
\end{axis}
\end{tikzpicture}
    \caption{Unitary fixed points for $N=8$ found using a numerical search beginning with 10,000 points.}
\label{fig:Neq8}
\end{figure}

We display the results of our numerical searches for $4\leq N\leq 8$ in Fig.\;\ref{fig:Neq4}\,-\ref{fig:Neq8}. The appearance of a leading-order conformal manifold of fixed points for the $O(4)$ and $B_4$ models inhibits our numerical searches, so these models are omitted from Fig.\;\ref{fig:Neq4}. It is important to note that none of the identified fixed points, ignoring the trivial fixed point, were stable against generic deformations. Of interest is the fact that the value of the $h_i^2$ invariant appears to be bounded from above for all of the interacting bulk theories, though we were unable to find an analytic form for such a bound. In all of the examined cases we find a rich zoo of symmetry breaking patterns. For the $O(N)$ models and free bulks the number of $\kappa=0$ eigenvalues gives an indication of the continuous symmetry group remains, though it does not allow us to determine if there are additional discrete group factors. We find that beyond the breaking to $O(N-1)$ studied previously, there are multiple breakings to smaller orthogonal groups in addition to a sea of fixed points with only discrete symmetry remaining. For the hypercubic and hypertetrahedral models it is possible to estimate the order of the symmetry group remaining at a fixed point using the number of physically equivalent theories. As the numerics inevitably do not find all of the fixed points in every orbit, this is an imperfect process, but indicates that for these groups, too, the space of interface dCFTs is rich in potential symmetry breaking patterns.

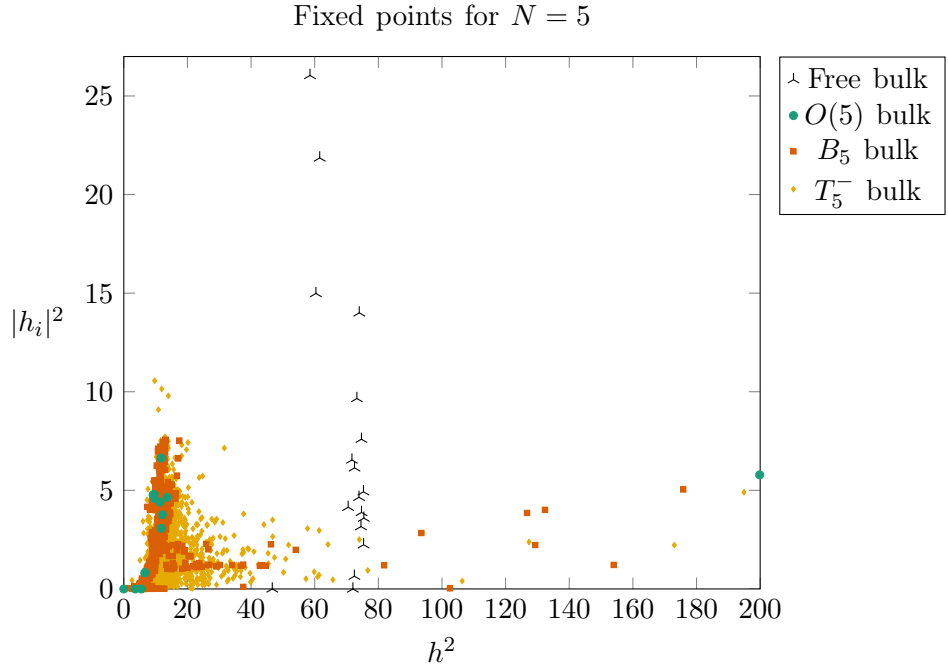
\begin{figure}[H]
\centering
\begin{tikzpicture}
\begin{axis}[
    xmin=-0, xmax=200,
    ymin=0, ymax=27,
    xlabel= $h^2$,
    ylabel= $|h_i|^2$,
    ylabel style={rotate=-90},
    title={Fixed points for $N=5$},
    legend pos = outer north east,
    reverse legend,
]

\addplot+[
    only marks,
    mark=diamond*,
    mark options={color=Dark2-F,fill=Dark2-F},
    mark size=1pt]
table{Data/T5m.dat};
\addplot+[
    only marks,
    mark=square*,
    mark options={color=Dark2-B,fill=Dark2-B},
    mark size=1pt]
table{Data/B5.dat};
\addplot+[
    only marks,
    mark=*,
    mark options={color=Dark2-A,fill=Dark2-A},
    mark size=1.5pt]
table{Data/O5.dat};
\addplot+[
    only marks,
    mark=Mercedes star,
    mark options={color=black},
    mark size=2.5pt]
table{Data/Free5.dat};
\legend{$T_5^-$ bulk, $B_5$ bulk, $O(5)$ bulk, Free bulk}
\end{axis}
\end{tikzpicture}
    \caption{Unitary fixed points for $N=5$ found using a numerical search beginning with 10,000 points.}
\label{fig:Neq5}
\end{figure}

\begin{figure}[H]
\centering
\begin{tikzpicture}
\begin{axis}[
    xmin=-0, xmax=200,
    ymin=0, ymax=27,
    xlabel= $h^2$,
    ylabel= $|h_i|^2$,
    ylabel style={rotate=-90},
    title={Fixed points for $N=6$},
    legend pos = outer north east,
    reverse legend,
]
\addplot+[
    only marks,
    mark=pentagon*,
    mark options={color=Dark2-H,fill=Dark2-H},
    mark size=1pt]
table{Data/T6p.dat};
\addplot+[
    only marks,
    mark=diamond*,
    mark options={color=Dark2-F,fill=Dark2-F},
    mark size=1pt]
table{Data/T6m.dat};
\addplot+[
    only marks,
    mark=square*,
    mark options={color=Dark2-B,fill=Dark2-B},
    mark size=1pt]
table{Data/B6.dat};
\addplot+[
    only marks,
    mark=*,
    mark options={color=Dark2-A,fill=Dark2-A},
    mark size=1.5pt]
table{Data/O6.dat};
\addplot+[
    only marks,
    mark=Mercedes star,
    mark options={color=black},
    mark size=2.5pt]
table{Data/Free6.dat};
\legend{$T_6^+$ bulk,$T_6^-$ bulk,$B_6$ bulk, $O(6)$ bulk, Free bulk}
\end{axis}
\end{tikzpicture}
    \caption{Unitary fixed points for $N=6$ found using a numerical search beginning with 10,000 points.}
\label{fig:Neq6}
\end{figure}
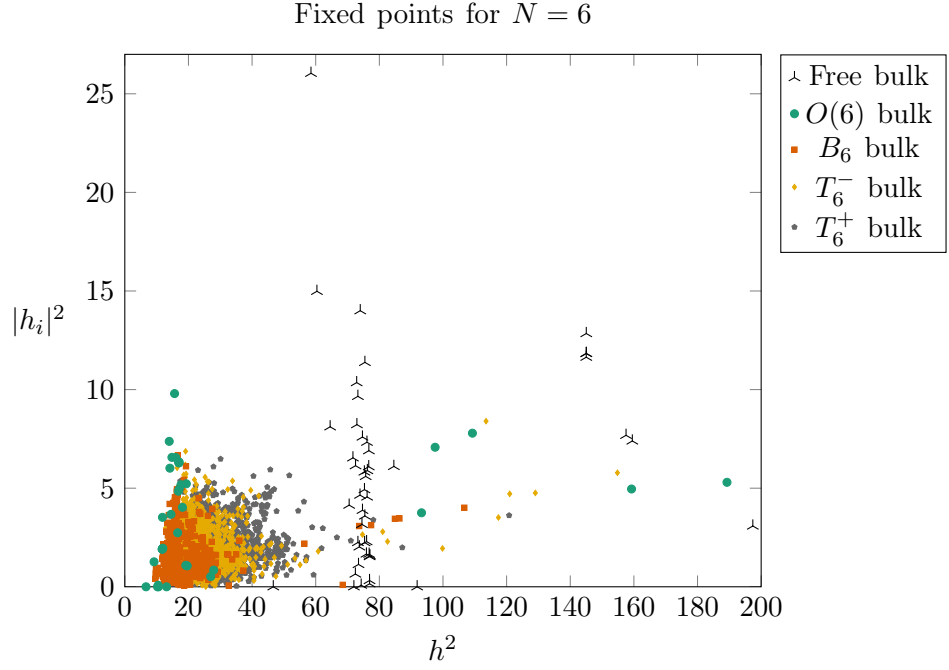

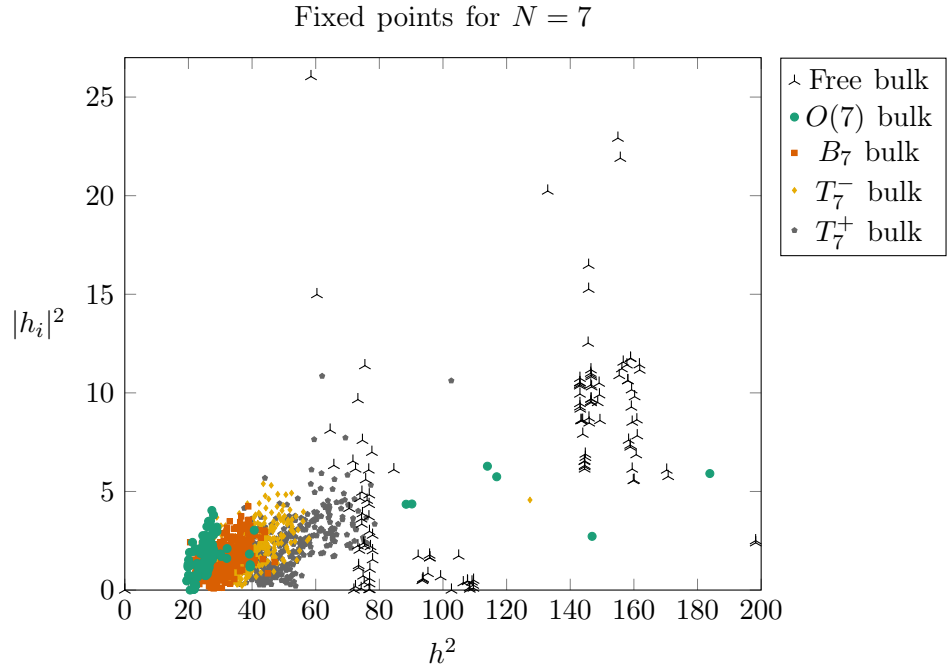
\begin{figure}[H]
\centering
\begin{tikzpicture}
\begin{axis}[
    xmin=-0, xmax=200,
    ymin=0, ymax=27,
    xlabel= $h^2$,
    ylabel= $|h_i|^2$,
    ylabel style={rotate=-90},
    title={Fixed points for $N=7$},
    legend pos = outer north east,
    reverse legend,
]
\addplot+[
    only marks,
    mark=pentagon*,
    mark options={color=Dark2-H,fill=Dark2-H},
    mark size=1pt]
table{Data/T7p.dat};
\addplot+[
    only marks,
    mark=diamond*,
    mark options={color=Dark2-F,fill=Dark2-F},
    mark size=1pt]
table{Data/T7m.dat};
\addplot+[
    only marks,
    mark=square*,
    mark options={color=Dark2-B,fill=Dark2-B},
    mark size=1pt]
table{Data/B7.dat};
\addplot+[
    only marks,
    mark=*,
    mark options={color=Dark2-A,fill=Dark2-A},
    mark size=1.5pt]
table{Data/O7.dat};
\addplot+[
    only marks,
    mark=Mercedes star,
    mark options={color=black},
    mark size=2.5pt]
table{Data/Free7.dat};
\legend{$T_7^+$ bulk,$T_7^-$ bulk,$B_7$ bulk, $O(7)$ bulk, Free bulk}
\end{axis}
\end{tikzpicture}
    \caption{Unitary fixed points for $N=7$ found using a numerical search beginning with 10,000 points.}
\label{fig:Neq7}
\end{figure}

\section{Conclusion}
In this work we studied the space of interface CFTs that arise when well-known CFTs in $4-\varepsilon$ dimensions are perturbed by localised interactions cubic in $\phi$. Our results indicate that this is an extensive space, comprising interface fixed points with global symmetries given by subgroups of the global symmetry of the bulk CFT in which they live. In many instances we find multiple interface CFTs with the same global symmetry, and we also find interface CFTs with no global symmetry whatsoever. The interface CFTs we have found always have at least one relevant operator. We have focused on unitary theories, but our methods are also applicable to non-unitary ones, which typically include fixed points that are RG stable as was already noticed in~\cite{Harribey:2023xyv}.

The study of fixed points within the framework of the perturbative $\varepsilon$ expansion is valuable, as it reveals aspects of the space of CFTs that can guide research with non-perturbative methods, e.g.\ the conformal bootstrap. While we have resorted to a standard perturbative treatment, analytic bootstrap methods can also be used to study the spectrum and operator product expansion coefficients of interface CFTs, as already discussed in \cite{Dey:2020jlc}. Of course our results here are strictly valid for $\varepsilon$ small, and their continuation to $\varepsilon=1$ is not obvious. Nevertheless, our results indicate that the space of two-dimensional interface CFTs that live in three-dimensional space is vast. Perhaps it can be charted by applying the bootstrap methods of \cite{Gliozzi:2015qsa} or extending those of \cite{Behan:2020nsf, Behan:2021tcn} to interfaces or by applying the folding trick. The theories with $SO(3)$, $SU(3)$, $Sp(6)$ and $F_4$ global symmetries discussed in this work would offer obvious targets.

Further possible extensions of this work include the consideration of surface defects in bulk universality classes beyond the $O(N)$ model, as has been done here for interfaces and in \cite{Pannell:2023pwz} for line defects. Composite defects~\cite{Shimamori:2024yms}, where the combined effects of an interface hosting a lower-dimensional defect can also be studied with the $\varepsilon$ expansion. This would presumably give rise to new universality classes that have not yet been explored. Multiple interfaces and the effective field theory that may be defined for them can also be analysed, using the framework recently discussed in~\cite{Diatlyk:2024qpr, Kravchuk:2024qoh}. Finally, the analysis of interfaces that separate two different CFTs in the $\varepsilon$ expansion would offer a variety of avenues for investigation.

\ack{We would like to thank Hugh Osborn for insightful comments that led to our discussion of the $F_4$ family of fixed points. The numerical computations in this work have used King's College London's CREATE~\cite{CREATE} computing resources. Some computations in this paper have been performed with the help of \emph{Mathematica} with the packages \href{http://www.xact.es}{\texttt{xAct}}~\cite{xact} and \href{http://www.xact.es/xTras/index.html}{\texttt{xTras}}~\cite{Nutma:2013zea}. Nordita is supported in part by NordForsk. AS is supported by the Royal Society under grant URF{\textbackslash}R1{\textbackslash}211417 and by STFC under grant ST/X000753/1. AS thanks the CERN Department of Theoretical Physics for hospitality during the final stages of completion of this work.}

\begin{appendix}

\section{Preliminary two-loop results}
\label{ap:2loops}

In this appendix we gather results on the computation of the two-loop interface beta function.

The beta function of the bulk coupling is not affected by the interface and is given by
\begin{align} \label{eq:beta4-general-2-loops}
		\beta_{ijkl} \, &= \, - \, \varepsilon \lambda_{ijkl}
		\, + \, \left(\lambda_{ijmn}\lambda_{mnkl} + 2 \textrm{ perms} \right)  \, - \, \left(\lambda_{ijmn}\lambda_{mefk}\lambda_{nefl}+ 5 \textrm{ perms} \right) \crcr
		&\quad + \, \frac{1}{12} \left(\lambda_{ijkm}\lambda_{mnef}\lambda_{nefl} + 3 \textrm{ perms} \right)  .
\end{align}

At two loops, three type of graphs contribute to the three-point function on the interface: one-particle-irreducible graphs with only cubic couplings, one-particle-irreducible graphs with both cubic and quartic couplings and one-particle-reducible graphs coming from corrections to the two-point function. They are represented in Fig.\;\ref{fig:2loopcubic}, \ref{fig:2loopmixed} and \ref{fig:2pt} respectively.

With the notation of \eqref{eq:3pointbare}, the three-point function at two loops is given by
\begin{align} \label{eq:bare_3pt_2loops}
 \mu^{-2\varepsilon}\mathcal{A}^{(2)}_{ijk} \,& = \,
 \big(h_{ide} h_{dfg } h_{efl } h_{jgm } h_{klm }+ 2 \textrm{ perms} \big) \,  I_1 + \big(h_{ide } h_{jdf} h_{kgf} h_{glm} h_{elm }+ 2 \textrm{ perms} \big) \,  I_2  \crcr
&\quad + h_{ide } h_{jfg } h_{klm} h_{dfl } h_{egm } \, I_3  + \big(\lambda_{ijef}\lambda_{efgl}h_{glk }+ 2 \textrm{ perms} \big) \,  B_2  \crcr
&\quad + \big( \lambda_{idef} \lambda_{jdel}h_{flk }+ 2 \textrm{ perms} \big) \,  S_1 + \big( \lambda_{ijef} \lambda_{eglk}h_{glf }+ 2 \textrm{ perms} \big) \,  S_2 \, , \crcr
&\quad +\big( \lambda_{iefg}h_{efl}h_{lmj } h_{gmk }+ 5 \textrm{ perms} \big) \,  U_1 +\big( \lambda_{ijef}h_{egl}h_{fgm}h_{mlk }+ 2 \textrm{ perms} \big) \,U_2  \crcr
&\quad +\big( \lambda_{efgl}h_{ief }h_{jgm } h_{klm}+ 2 \textrm{ perms} \big) \,  U_3 +\big( \lambda_{ijde}h_{kdf}h_{fml }h_{eml }+ 2 \textrm{ perms} \big) \,  U_4
\crcr
&\quad +\big( \lambda_{kdml}h_{ide }h_{jfl } h_{efm }+ 2 \textrm{ perms} \big) \,  U_5  +\big( \lambda_{efgl} \lambda_{kfgl}h_{ije }+ 2 \textrm{ perms} \big) M \, \crcr
&\quad + \big(h_{ije}h_{efg}h_{lmk} \lambda_{fglm}+ 2 \textrm{ perms} \big)K_1+ \big(h_{ije}h_{fgl}h_{glm} \lambda_{efmk}+ 2 \textrm{ perms} \big) K_2\crcr
&\quad + \big(h_{ije}h_{efg}h_{flm} \lambda_{lmgk}+ 2 \textrm{ perms} \big)K_3 +\big(h_{ije}h_{lmg}h_{gfk} \lambda_{elmf}+ 2 \textrm{ perms} \big)K_4\, .
\end{align}

We now need to compute the counterterms needed to make the three-point function finite. They can be obtained directly for each graph by using the Bogoliubov-Parasuk recursion formula \cite{Bogoliubov:1957gp}. We will then project onto the simple poles in $1/\varepsilon$ to obtain the beta function coefficients.

Denoting
\begin{align}
    Z^{(1)}_{ijk}&= a_T h_{ide }h_{jdf}h_{kef}
+ a_{I_1}\big(h_{ide} h_{dfg } h_{efl } h_{jgm } h_{klm }+ 2 \textrm{ perms} \big)  \crcr
&
\quad+ a_{I_2}\big(h_{ide } h_{jdf} h_{kgf} h_{glm} h_{elm }+ 2 \textrm{ perms} \big) \,
+ a_{I_3}h_{ide } h_{jfg } h_{klm} h_{dfl } h_{egm } \crcr
&\quad+a_{B}\big(\lambda_{ijef}h_{efk }+ 2 \textrm{ perms} \big) \,  + a_{B_2}\big(\lambda_{ijef}\lambda_{efgl}h_{glk }+ 2 \textrm{ perms} \big)   \crcr
&\quad + a_{S_1}\big( \lambda_{idef} \lambda_{jdel}h_{flk }+ 2 \textrm{ perms} \big) \,  + a_{S_2}\big( \lambda_{ijef} \lambda_{eglk}h_{glf }+ 2 \textrm{ perms} \big) \,  \, , \crcr
&\quad +a_{U_1}\big( \lambda_{iefg}h_{efl}h_{lmj } h_{gmk }+ 5 \textrm{ perms} \big) +a_{U_2}\big( \lambda_{ijef}h_{egl}h_{fgm}h_{mlk }+ 2 \textrm{ perms} \big) \,  \crcr
&\quad +a_{U_3}\big( \lambda_{efgl}h_{ief }h_{jgm } h_{klm}+ 2 \textrm{ perms} \big)+a_{U_4}\big( \lambda_{ijde}h_{kdf}h_{fml }h_{eml }+ 2 \textrm{ perms} \big) \,
\crcr
&\quad +a_{U_5}\big( \lambda_{kdml}h_{ide }h_{jfl } h_{efm }+ 2 \textrm{ perms} \big) \,  + a_{M}\big( \lambda_{efgl} \lambda_{kfgl}h_{ije }+ 2 \textrm{ perms} \big) \, \crcr
&\quad +a_{K_1} \big(h_{ije}h_{efg}h_{lmk} \lambda_{fglm}+ 2 \textrm{ perms} \big)+ a_{K_2}\big(h_{ije}h_{fgl}h_{glm} \lambda_{efmk}+ 2 \textrm{ perms} \big) \crcr
&\quad + a_{K_3}\big(h_{ije}h_{efg}h_{flm} \lambda_{lmgk}+ 2 \textrm{ perms} \big)+ a_{K_4}\big(h_{ije}h_{lmg}h_{gfk} \lambda_{elmf}+ 2 \textrm{ perms} \big)\, ,
\end{align}
we have
\begin{align}
 a_{T} \, & =  \, - (2\pi)^{2} \, T^{(1)} \, , &
a_{I_1} \, &= \,  (2\pi)^{4} \, (-I_1^{(1)}+T^{(1)}T^{(0)}) \, , \crcr
a_{I_2} \, & =\, -  (2\pi)^{4} \, I_2^{(1)}\, , &
a_{I_3} \,  &=\, -  (2\pi)^{4} \, I_3^{(1)} \, , \crcr
a_B&=-(4\pi)^{2} B^{(1)} \, , &
a_{S_1} \, &= (4\pi)^{4}(2B^{(0)}D^{(1)}-S_1^{(1)})\crcr
a_{S_2} \, &= (4\pi)^{4}\left(2B^{(0)}B^{(1)}-S_2^{(1)}\right)\, , &
a_{B_2} &= (4\pi)^{4}\left(B^{(0)}B^{(1)}+B^{(0)}D^{(1)}-B_2^{(1)}\right) \, , \crcr
a_{U_1}&=4(2\pi)^{4}\left(B^{(1)}T^{(0)}-U_1^{(1)}\right) \, , & a_{U_2} \, &=  4(2\pi)^{4}\left(B^{(0)}T^{(1)}-U_2^{(1)}\right) \, ,   \crcr
a_{U_3} \, &=  4(2\pi)^{4}\left(B^{(1)}T^{(0)}-U_3^{(1)} \right)\, , & a_{U_4} \, &= -4(2\pi)^{4}U_4^{(1)} \, , \crcr
a_{U_5} \, &=  -4(2\pi)^{4}U_5^{(1)} \, , & a_{K_1} \, &= -4(2\pi)^{4}K_1^{(1)} \, , \crcr
a_{K_2} \, &=  -4(2\pi)^{4}K_2^{(1)} \, , & a_{K_3}&=a_{K_4} \, = -4(2\pi)^{4}K_3^{(1)} \, , \crcr
a_{M} \, &=  -(4\pi)^{4}\left(M^{(1)}+\tfrac{1}{24}\right) \, , & &
\label{eq:ct2loops}
\end{align}
where we have parameterised the amplitude of a $\ell$-loop graph $\mathcal{G}$ as $\mathcal{A}(\mathcal{G})=\mu^{-\ell\varepsilon}\sum_{k=0}^\ell \frac{\mathcal{G}^{(k)}}{\varepsilon^k}$. $D$ is the amplitude of the bulk one-loop bubble graph with two quartic couplings contributing to the renormalisation of the bulk coupling
\begin{equation}\label{eq:Dcomp}
    D=\begin{tikzpicture}[baseline=(vert_cent.base), square/.style={regular polygon,regular polygon sides=4}]
        \node at (0,0) [square,draw,fill=white,inner sep=1.2pt,outer sep=0pt]  (l) {};
        \node at (1.2,0) [square,draw,fill=white,inner sep=1.2pt,outer sep=0pt] (r) {};
        \draw[densely dashed] (l) to [out=30,in=150] (r);
        \draw[densely dashed] (l) to [out=-30,in=-150] (r);
        \draw[densely dashed] (l)--++(150:0.4cm);
        \draw[densely dashed] (l)--++(-150:0.4cm);
        \draw[densely dashed] (r)--++(30:0.4cm);
        \draw[densely dashed] (r)--++(-30:0.4cm);
        \node[inner sep=0pt,outer sep=0pt] (vert_cent) at (0,0) {$\phantom{\cdot}$};
    \end{tikzpicture}
    =-\frac{\mu^{-\varepsilon}}{16\pi^2\varepsilon} +\text{O}(\varepsilon^0) \, .
\end{equation}
The counterterm in the coefficient $a_M$ comes from the one-loop part of the bulk wavefunction renormalisation factor \eqref{eq:wfbulk}.

Using \eqref{eq:genericbetaMS}, we finally obtain the two-loop beta functions of the interface coupling
\begin{align}
    \beta_{ijk}&=-\frac{\varepsilon}{2}h_{ijk}+a_T h_{ide }h_{jdf}h_{kef}
+ 2\lsp a_{I_1}\big(h_{ide} h_{dfg } h_{efl } h_{jgm } h_{klm }+ 2 \textrm{ perms} \big)  \crcr
&
+ 2\lsp a_{I_2}\big(h_{ide } h_{jdf} h_{kgf} h_{glm} h_{elm }+ 2 \textrm{ perms} \big) \,
+ 2\lsp a_{I_3}h_{ide } h_{jfg } h_{klm} h_{dfl } h_{egm } \crcr
&+a_{B}\big(\lambda_{ijef}h_{efk }+ 2 \textrm{ perms} \big) \,  + 2\lsp a_{B_2}\big(\lambda_{ijef}\lambda_{efgl}h_{glk }+ 2 \textrm{ perms} \big)   \crcr
& + 2\lsp a_{S_1}\big( \lambda_{idef} \lambda_{jdel}h_{flk }+ 2 \textrm{ perms} \big) \,  + 2\lsp a_{S_2}\big( \lambda_{ijef} \lambda_{eglk}h_{glf }+ 2 \textrm{ perms} \big) \,  \, , \crcr
& +2\lsp a_{U_1}\big( \lambda_{iefg}h_{efl}h_{lmj } h_{gmk }+ 5 \textrm{ perms} \big) +2\lsp a_{U_2}\big( \lambda_{ijef}h_{egl}h_{fgm}h_{mlk }+ 2 \textrm{ perms} \big) \,  \crcr
& +2\lsp a_{U_3}\big( \lambda_{efgl}h_{ief }h_{jgm } h_{klm}+ 2 \textrm{ perms} \big)+2\lsp a_{U_4}\big( \lambda_{ijde}h_{kdf}h_{fml }h_{eml }+ 2 \textrm{ perms} \big) \,
\crcr
& +2\lsp a_{U_5}\big( \lambda_{kdml}h_{ide }h_{jfl } h_{efm }+ 2 \textrm{ perms} \big) \,  + 2\lsp a_{M}\big( \lambda_{efgl} \lambda_{kfgl}h_{ije }+ 2 \textrm{ perms} \big) \, \crcr
& +2\lsp a_{K_1} \big(h_{ije}h_{efg}h_{lmk} \lambda_{fglm}+ 2 \textrm{ perms} \big)+ 2\lsp a_{K_2}\big(h_{ije}h_{fgl}h_{glm} \lambda_{efmk}+ 2 \textrm{ perms} \big) \crcr
& + 2\lsp a_{K_3}\big(h_{ije}h_{efg}h_{flm} \lambda_{lmgk}+ 2 \textrm{ perms} \big)+ 2\lsp a_{K_4}\big(h_{ije}h_{lmg}h_{gfk} \lambda_{elmf}+ 2 \textrm{ perms} \big)\, .
\label{eq:betatwo}
\end{align}

The amplitudes of the pure interface graphs can be computed and our results were cross-checked with the long-range computation of \cite{Theumann:1985qc}.
The computation of the mixed graphs is more challenging due to the interface-to-bulk propagator and we were not able to compute all the beta function coefficients. We have
\begin{align}
    a_T&=-\frac{1}{4} \, ,\quad  a_B=1 \, , \crcr
    2\lsp a_{I_1}&=-\frac{1}{16}  \, ,\quad 2\lsp a_{I_2}= \frac{1}{64} \, ,\quad 2\lsp a_{I_3}=-\frac{\pi^2}{128} \, , \crcr
    2\lsp a_{B_2}&=0 \, , \quad 2\lsp a_{S_1}=-1 \, , \quad 2\lsp a_{S_2}=-1 \, , \crcr
    2\lsp a_{U_2}&=\frac{1}{4} \, \quad 2\lsp a_{U_3}=0 \, \quad 2\lsp a_{U_4}=-\frac{1}{8}  \, , \crcr
   2\lsp a_{K_2}&=\frac{1}{16}  \, , \quad 2\lsp a_{M}= \frac{1}{12} \, .
\end{align}

We still have to compute $a_{U_1},a_{U_5},a_{K_1}$ and $a_{K_3}$. In the next subsection we give more details on the computation of the bulk-defect integrals that we have computed so far.

\subsection{Computation of bulk-defect integrals}

To compute the amplitudes of the graphs contributing to the three-point function we will use repeatedly the following two formulas
\begin{equation}
\label{eq:bubble}
\int \frac{d^{\lsp d} \vec{q}}{(2\pi)^d}\frac{1}{|\vec{q}|^{2\alpha} |\vec{q+p}|^{2\beta}}=\frac{|\vec{p}|^{d-2\alpha-2\beta}}{(4\pi)^{d/2}}\frac{\Gamma(\frac{d}{2}-\alpha)\Gamma(\frac{d}{2}-\beta)\Gamma(\alpha+\beta-\frac{d}{2})}{\Gamma(\alpha)\Gamma(\beta)\Gamma(d-\alpha-\beta)} \,,
\end{equation}
and
\begin{equation}
    \int \frac{d^{\lsp d} \vec{q}}{(2\pi)^d} \frac{1}{|\vec{q}|^a(|\vec{q}|+|\vec{p}|)}=\frac{2|\vec{p}|^{d-1-a}}{\Gamma(\frac{d}{2})(4\pi)^{d/2}}\Gamma(d-a)\Gamma(1-d+a) \, .
    \label{eq:simpbubble}
\end{equation}

\subsubsection{Graphs with only one interface coupling}

To compute these integrals we use the trick explained at the end of section \ref{app:MS}. We obtain
\begin{equation}
    2\lsp a_{S_1}=2a_{S_2}=-1 \, ,\qquad 2\lsp a_{B_2}=0 \, , \qquad 2\lsp a_M=\frac{1}{12} \, .
\end{equation}

We were able to compute independently the amplitude $B_2$, providing a cross-check of this method.

We have
\begin{align}
    B_2&=\frac{1}{64} \int \frac{d^{\lsp d} \vec{q_1} d^{\lsp d} \vec{q_2}}{(2\pi)^{2d}}\int dy \int dz \frac{e^{-2|\vec{q_1}||z|}e^{-2|\vec{q_2}||z-y|}e^{-2|\vec{p}||y|}}{|\vec{q_1}|^2|\vec{q_2}|^2}  \crcr
    &= \frac{1}{64} \int \frac{d^{\lsp d} \vec{q_1} d^{\lsp d} \vec{q_2}}{(2\pi)^{2d}}\frac{1}{|\vec{q_1}|^2|\vec{q_2}|^2} \int dz\, e^{-2|\vec{q_1}||z|}\frac{|\vec{q_2}|e^{-2|\vec{p}||z|}-|\vec{p}|e^{-2|\vec{q_2}||z|}}{(|\vec{q_2}|+|\vec{p}|)(|\vec{q_2}|-|\vec{p}|)}\crcr
    &= \frac{1}{64} \int \frac{d^{\lsp d} \vec{q_1} d^{\lsp d} \vec{q_2}}{(2\pi)^{2d}}\frac{1}{|\vec{q_1}|^2|\vec{q_2}|^2(|\vec{q_2}|+|\vec{p}|)(|\vec{q_2}|-|\vec{p}|)} \left(\frac{|\vec{q_2}|}{|\vec{p}|+|\vec{q_1}|}-\frac{|\vec{p}|}{|\vec{q_1}|+|\vec{q_2}|}\right) .
\end{align}

It is then possible to integrate over $\vec{q_1}$ using \eqref{eq:simpbubble} to obtain
\begin{align}
    B_2&=\frac{\mu^{-2\varepsilon}}{16\lsp\Gamma^2(\frac{d}{2})(4\pi)^d}\Gamma(d-2)\Gamma(3-d)\int dq\, \frac{q^{d-2}-q^{2d-6}}{(q+1)(q-1)} \crcr
    &=-\frac{\mu^{-2\varepsilon}}{32\lsp\Gamma(\frac{d}{2})^2(4\pi)^d}\Gamma(d-2)\Gamma(3-d) \frac{\pi\tan(\frac{\pi d}{2})}{\cos(\pi d)} \crcr
    &= \frac{\mu^{-2\varepsilon}}{256 \pi^4 }\left( \frac{1}{\varepsilon^2}+\frac{2-\gamma+\ln\pi}{ \varepsilon} +\text{O}(\varepsilon^0)\right) .
\end{align}
Using \eqref{eq:ct2loops}, we then indeed obtain $a_{B_2}=0$.

\subsubsection{Graphs with more than one interface coupling}

\paragraph{The integral $\boldsymbol{U_2}$}

We have
\begin{align}
    U_2&=-\frac{1}{64}\int \frac{d^{\lsp d} \vec{q_1} d^{\lsp d} \vec{q_2}}{(2\pi)^{2d}}\int dy \frac{e^{-2(|\vec{p}|+|\vec{q_1}|)|y|}}{|\vec{q_1}|^2|\vec{q_2}||\vec{q_1}+\vec{q_2}|^2}\crcr
    &=-\frac{1}{64}\int \frac{d^{\lsp d} \vec{q_1} d^{\lsp d} \vec{q_2}}{(2\pi)^{2d}}\frac{1}{|\vec{q_1}|^2|\vec{q_2}||\vec{q_1}+\vec{q_2}|^2(|\vec{p}|+|\vec{q_1}|)} \, .
\end{align}
We use \eqref{eq:bubble} to integrate over $\vec{q_2}$ and
\eqref{eq:simpbubble} to integrate over $\vec{q_1}$. We obtain
\begin{align}
    U_2&=\frac{\mu^{-2\varepsilon}}{32(4\pi)^d}\frac{\Gamma(\frac{d-1}{2})\Gamma(\frac{d-2}{2})\Gamma(\frac{3-d}{2})\Gamma(1-2d)\Gamma(2d)}{\Gamma(\frac{1}{2})\Gamma(d-\frac{3}{2})\Gamma(\frac{d}{2})} \crcr
    &= -\frac{\mu^{-2\varepsilon}}{512\pi^4}\left( \frac{1}{ \varepsilon^2}+\frac{3-\gamma+\ln\pi}{ \varepsilon}+\mathcal{O}(\varepsilon^0)  \right) .
\end{align}
For the corresponding beta function coefficient we find
$2\lsp a_{U_2}=\frac{1}{4}$.

\paragraph{The integral $\boldsymbol{U_4}$}

We have
\begin{align}
    U_4&=-\frac{1}{64}\int \frac{d^{\lsp d} \vec{q_1} d^{\lsp d} \vec{q_2}}{(2\pi)^{2d}} \int dy \frac{e^{-2(|\vec{q_1}|+|\vec{p}|)|y|}}{|\vec{q_1}|^3|\vec{q_2}||\vec{q_1}+\vec{q_2}|}\crcr
    &=-\frac{1}{64}\int \frac{d^{\lsp d} \vec{q_1} d^{\lsp d} \vec{q_2}}{(2\pi)^{2d}} \frac{1}{|\vec{q_1}|^3|\vec{q_2}||\vec{q_1}+\vec{q_2}|(|\vec{q_1}|+|\vec{p}|)} \crcr
    & =\frac{\mu^{-2\varepsilon}}{32(4\pi)^d}\frac{\Gamma(\frac{d-1}{2})^2\Gamma(\frac{2-d}{2})\Gamma(2d)\Gamma(1-2d)}{\Gamma(\frac{d}{2})\Gamma(\frac{1}{2})^2\Gamma(d-1)} \crcr
    & = \frac{\mu^{-2\varepsilon}}{1024\pi^4 \varepsilon}+\text{O}(\varepsilon^0) \, ,
\end{align}
where we first integrated over $\vec{q_2}$ using \eqref{eq:bubble} and then
over $\vec{q_1}$ using \eqref{eq:simpbubble}. For the corresponding beta
function coefficient we then obtain $2\lsp a_{U_4}=-\frac{1}{8}$.

\paragraph{The integral $\boldsymbol{K_2}$}

We have
\begin{align}
    K_2&=-\frac{1}{128|\vec{p}|}\int \frac{d^{\lsp d} \vec{q_1} d^{\lsp d} \vec{q_2}}{(2\pi)^{2d}} \int dy \frac{e^{-2(|\vec{q_1}|+|\vec{p}|)|y|}}{|\vec{q_1}|^2|\vec{q_2}||\vec{q_1}+\vec{q_2}|}\crcr
    &=-\frac{1}{128|\vec{p}|}\int \frac{d^{\lsp d} \vec{q_1} d^{\lsp d} \vec{q_2}}{(2\pi)^{2d}} \frac{1}{|\vec{q_1}|^2|\vec{q_2}||\vec{q_1}+\vec{q_2}|(|\vec{q_1}|+|\vec{p}|)} \crcr
    & =-\frac{\mu^{-2\varepsilon}}{128(4\pi)^d}\frac{\Gamma(\frac{d-1}{2})^2\Gamma(\frac{2-d}{2})\Gamma(2d-4)\Gamma(5-2d)}{\Gamma(\frac{d}{2})\Gamma(\frac{1}{2})^2\Gamma(d-1)} \crcr
    & =-\frac{\mu^{-2\varepsilon}}{4096\pi^4 \varepsilon}+\text{O}(\varepsilon^0) \, ,
\end{align}
where we first integrated over $\vec{q_2}$ using \eqref{eq:bubble} and then
over $\vec{q_1}$ using \eqref{eq:simpbubble}. For the corresponding beta
function coefficient we obtain $2\lsp a_{K_2}=\frac{1}{16}$.

\end{appendix}

\bibliography{Refs}

\end{document}